\documentclass[pra, notitlepage,  amsmath,amssymb,superscriptaddress,nofootinbib]{revtex4-1}

\usepackage{setspace}

\usepackage{amsmath}
\usepackage{amsfonts}            
\usepackage{amssymb}

\usepackage{mathrsfs}

\usepackage{eufrak}
\usepackage{color}

\usepackage{graphicx}
\usepackage{epstopdf}

\usepackage{dcolumn}
\usepackage{bm}

\makeatletter
\def\@dotsep{4.5}
\makeatother

\begin{document}

\title{Manipulation of Molecules with Electromagnetic Fields}

\author{Mikhail Lemeshko}
\email{mikhail.lemeshko@gmail.com}
\affiliation{ITAMP, Harvard-Smithsonian Center for Astrophysics, 60 Garden Street, Cambridge, MA 02138, USA}%
\affiliation{Physics Department, Harvard University, 17 Oxford Street, Cambridge, MA 02138, USA} %
\affiliation{Kavli Institute for Theoretical Physics, University of California, Santa Barbara, CA 93106, USA}

\author{Roman V. Krems}
\email{rkrems@chem.ubc.ca}
\affiliation{Department of Chemistry, University of British Columbia, Vancouver, BC V6T 1Z1, Canada }
\affiliation{Kavli Institute for Theoretical Physics, University of California, Santa Barbara, CA 93106, USA}

\author{John M. Doyle}
\affiliation{Physics Department, Harvard University, 17 Oxford Street, Cambridge, MA 02138, USA} %

\author{Sabre Kais}
\affiliation{Departments of Chemistry and Physics, Purdue University, West Lafayette, Indiana 47907, USA} %

\date{\today}

\begin{abstract}



\end{abstract}


\maketitle

\section{Introduction}
\label{intro}

The past few decades have witnessed remarkable developments of laser techniques setting the stage for new areas of research in molecular physics. It is now possible to interrogate molecules in the ultrafast and ultracold regimes of molecular dynamics and the measurements of molecular structure and dynamics can be made with unprecedented precision. 
 Molecules are inherently complex quantum-mechanical systems. The complexity of molecular structure, if harnessed, can be exploited for yet another step forward in science, potentially leading to technology for quantum computing, quantum simulation, precise field sensors, and new lasers.

 The goal of the present article is to review the major developments that have led to the current understanding of molecule - field interactions and experimental methods for manipulating molecules with electromagnetic fields. Molecule - field interactions are at the core of several, seemingly distinct,  areas of molecular physics. This is reflected in the organization of this article, which includes sections on {\it Field control of molecular beams}, {\it External field traps for cold molecules}, {\it Control of molecular orientation and molecular alignment}, {\it Manipulation of molecules by non-conservative forces}, {\it Ultracold molecules and ultracold chemistry}, {\it Controlled many-body phenomena}, {\it Entanglement of molecules and dipole arrays}, and {\it Stability of molecular systems in high-frequency super-intense laser fields}. By combining these topics in the same review, we would like to emphasize that all this work is based on the same basic Hamiltonian.  

This review is also intended to serve as an introduction to the excellent collection of articles appearing in this same-titled volume of {\it Molecular Physics}~\cite{HallMolPhys13, ManzMolPhys13, HermanMolPhys13, SinghMolPhys13, AverbukhMolPhys13, KotochigovaMolPhys13, TrippelMolPhys13, GuoMolPhys13, WeimerMolPhys13, BrouardMolPhys13, SchollkopfMolPhys13, TomzaMolPhys13, StuhlMolPhys13, DeMilleMolPhys13, PananghatMolPhys13, SpielerMolPhys13, WeiMolPhys13, ManaiMolPhys13, MerzMolPhys13, GarttnerMolPhys13, ToenniesMolPhys13, MullinMolPhys13, SadeghpourMolPhys13, GorshkovMolPhys13, LinMolPhys13, JansenMolPhys13, BohnMolPhys13}. These original contributions demonstrate the latest developments exploiting control of molecules with electromagnetic fields. The reader will be treated to a colourful selection of articles on topics as diverse as {\it Chemistry in laser fields}, {\it Quantum dynamics in helium droplets}, {\it Effects of microwave and laser fields on molecular motion}, {\it Rydberg molecules}, {\it Molecular structure in external fields}, {\it Quantum simulation with ultracold molecules}, and {\it Controlled molecular interactions},  written by many of the leading protagonists of these fields. 

This article is concerned chiefly with the effects of electromagnetic fields on low-energy rotational, fine-structure and translational degrees of freedom. There are several important research areas that are left outside the scope of this paper, most notably the large body of work on the interaction of molecules with attosecond laser pulses and high harmonic generation \cite{CorkumNatPhys07}, coherent control of molecular dynamics \cite{ShapiroRPP03} and optimal control of molecular processes \cite{BrifACP12}.  
We limit the discussion of resonant interaction of light with molecules to laser cooling strategies. We do not survey spectroscopy or transfer of population between molecular states. Even with these restrictions, this is a vast area to review, as is apparent from the number of references. 
We did our best to include all relevant results, with an emphasis on experimental work. However, some references may have been inadvertently omitted, for which the authors apologize. The in-depth discussion of selected topics of this article can be found in previous review articles. In particular, Refs.~\cite{DoyleEPJD04, CarrNJP09, FriedrichCPC09, SchnellAngChem09, BellMolPhys09, KreStwFrieColdMol, NiPCCP09, JinYePhysToday11, JinCREv12, QuemenerCrev12, BaranovCRev12} provide an introduction to the field of cold molecules, Refs.~\cite{vandeMeerakkerNature08, vandeMeerakkerChemRev12, BethlemIRPC03, DulieuRepProgPhys09, FriedrichCPC09, HoganPCCP11} review the work on controlled molecular beams, and Refs. ~\cite{StapelfeldtSeidemanRMP03, StapelfeldtPhysScr04, KumarappanPhysScr07, OhshimaIRPC10, SakaiBook} discuss molecules in short laser pulses.

This special issue and review article have been prepared in celebration of the 60th birthday of Professor Bretislav Friedrich, who is currently a Research Group Leader at the Fritz Haber Institute of the Max Planck Society and Honorarprofessor at the Technische Universit\"{a}t Berlin. Bretislav has made key contributions that became the origin of several of the research areas described here. 
Bretislav's work is an example of transformative research that converts simple ideas into profound results. Inspired by Bretislav's work is the present article.

\section{Molecules in fields}



\subsection{Individual molecules}

Unlike atoms, molecules possess the vibrational and rotational degrees of freedom, with characteristic energy scales $\sim$100~MHz -- 100 THz. Electronic excitations in molecules usually correspond to optical transition frequencies ($10^{14} - 10^{15}$~Hz), vibrational excitations to the infrared region ($10^{13} - 10^{14}$~Hz), and rotational excitations to the microwave region ($10^9 - 10^{11}$~Hz). In addition to the spin-orbit and hyperfine interactions, intrinsic to any atomic or molecular system, the rotational motion of molecules gives rise to new perturbations, 
such as $\Lambda$-doubling,  which generates closely spaced energy levels with characteristic excitation frequencies $< 100$ MHz~\cite{LevebvreBrionField2}. The eigenstates of a molecular Hamiltonian can be generally expanded in products of the electronic, vibrational and rotational states dressed by the electron and nuclear spin states. The interactions of the electrons and nuclei with the static and far-off-resonant optical fields used in experiments give rise to matrix elements with magnitudes $\leq $ 100 GHz.  Except in special cases, these interactions are unlikely to perturb the electronic and vibrational structure of molecules. On the other hand, the rotational, fine and hyperfine structure can be strongly altered by an applied field. In this section, we summarize general considerations at the core of the analysis of molecular structure in dc and non-resonant ac fields. Sec.~\ref{sec:cooling} discusses the effects of resonant and nearly resonant optical fields on molecular energy levels. 

In the absence of external fields, the molecular energy levels can be labeled by four quantum numbers, corresponding to the eigenvalues of four commuting operators: the Hamiltonian, the square of the total angular momentum $\bm{J}^2$, the $Z$-component of $\bm{J}$ and the parity operator, in addition to other (almost) good quantum numbers, if any. The presence of an axially symmetric external field perturbs the isotropy of space, which leads to couplings between states of different $J$. The operators describing field-induced interactions can be expressed in terms of rank$-1$ spherical tensors, leading to selection rules $|\Delta J| = 0, 1$ for couplings in first order and $|\Delta J | = 0, 2$ for couplings in second order. 
 Since the rotation about the field axis leaves the system invariant, the projection $M$ of the total angular momentum on the field direction remains a good quantum number. This quantum number is no longer good if two non-parallel fields are applied. 

Maxwell's equations define the magnetic $\bm{B}$ and electric $\bm{E}$ fields in terms of the vector $\bm{A}$ and scalar $\varphi$ potentials as follows 
\begin{eqnarray}
\bm{B} = \nabla \times \bm{A} \\
\bm{E} = \nabla \varphi - \frac{\partial \bm{A}}{\partial t}.
\end{eqnarray}
This shows that $\bm{E}$ is a vector, whereas $\bm{B}$ is a pseudovector.  As a result, electric-field-induced interactions must couple states of different parity, while interactions induced by the magnetic field must conserve parity. It should also be noted  that, for an electromagnetic wave, the magnitudes of the magnetic and electric fields are related as $E = Bc$, where $c$ is the speed of light. This implies that it is much easier to perturb the molecular energy levels by an electric field  than by a magnetic one. 

  The effect of a dc electric field on the molecular structure is determined by the matrix elements of the operator 
\begin{equation}
\label{StarkHam}
V_E = - \boldsymbol{\mu}  \mathbf{\cdot E},
\end{equation}
where  $\boldsymbol{\mu} = \sum_n q_n \mathbf{r}_n $, the vectors $\mathbf{r}_n$ give the positions of charges $q_n$ and the sum extends over both the electrons and the nuclei in the molecule. For molecules in a particular electronic and vibrational state, it is convenient to define the permanent dipole operator $\bm{d} = \langle \psi | \bm{\mu}| \psi \rangle$, where $|\psi \rangle$ is a product of the electronic and vibrational states. This vector operator depends on the angles that specify the orientation of the molecule's symmetry axis in a space-fixed coordinate frame. In the ground electronic state, the magnitude of $\bm{d}$ ranges from zero for nonpolar molecules to $\sim 10$ Debye for strongly polar molecules.

The interaction of a molecule with an electric field can be characterized by a dimensionless parameter $\eta = d E/\Delta_\pm$, which gives the ratio of the Stark energy to the splitting of the opposite parity levels $\Delta_\pm$ at zero field. For example, for closed-shell linear molecules, $\Delta_\pm$ is determined by the rotational constant $B_e$, while for symmetric top molecules or open-shell molecules in a $\Pi$ electronic state $\Delta_\pm$ is given by the value of the $\Lambda$-splitting.
The Stark shifts of the molecular energy levels are approximately quadratic functions of the electric field magnitude when $\eta \ll 1$ and linear when $\eta \gg 1$. 

In the case of a linear rigid rotor, an electric field creates a coherent superposition of the rotational states
\begin{equation}
\label{WFRigRot}
	\vert \tilde{J}M \rangle = \sum_J c_{J}^{\tilde{J}M} \vert J M \rangle.
\end{equation}
While $J$ is not a good quantum number, it is often possible to introduce an adiabatic label, $\tilde{J}$, such that the hybrid state $\vert \tilde{J}M \rangle \to \vert J M \rangle$ when the electric field magnitude $E \to 0$. The degree of molecular orientation in the laboratory frame is given by the expectation value of the orientation cosine, $\langle \cos \theta \rangle_{\tilde{J}M} = \langle \tilde{J}M \vert  \cos\theta \vert \tilde{J}M \rangle$, which corresponds to the expectation value of the electric dipole moment  in the space-fixed frame, $\langle d \rangle_{\tilde{J}M} = d \langle \cos \theta \rangle_{\tilde{J}M}$. Note that at any finite value of the electric field, the molecular orientation is confined to a finite range of angles, e.g.\ at a relatively strong field value of 100~kV/cm the axis of a NaK molecule ($d=2.7$~Debye, $B_e = 0.091$~cm$^{-1}$) keeps librating with the amplitude $\pm 25^\circ$. According to the uncertainty principle, in order to achieve a perfect orientation, $\langle \cos \theta \rangle = 1$, one would need to hybridize an infinite number of the angular momentum states.

A force acting on a molecule in an electric field is given by the negative gradient of the Stark energy, $W_\text{Stark}$,
\begin{equation}
\label{StarkForce}
	\mathbf{F} = - \nabla W_\text{Stark} (E)
\end{equation}
This can be used to confine molecular ensembles in external field traps and manipulate (accelerate, decelerate or deflect) molecular beams. 
The ground state of a molecule in an external field is always high-field seeking. Since creating a local maximum of a dc field is forbidden by the Earnshaw theorem~\cite{Earnshaw1839, WingPRL80}, trapping and decelerating high-field-seeking molecules presents a challenge. A number of techniques such as Alternating-Gradient Stark Deceleration~\cite{AuerbachJCP66, BethlemPRL02, TarbuttPRL04, BethlemJPB06, WohlfartPRA08, WohlfartPRA08b, KupperFarDiss09, KalninsRSI02}, ac electric trapping~\cite{vanVeldhovenPRL05, BethlemPRA06, SchnellJPCA07, LutzowPRA08}, and and optical deceleration and trapping~\cite{FriHerPRL95, MaherNatPhot12, FultonNatPhys06, FultonPRL04, BarkerPRA02}  have been developed to deal with this problem, as discussed in Sec.~\ref{sec:beams}. Due to a high density of states, all states of complex polyatomic molecules are high-field seeking. Trapping weak-field-seekers in a field minimum is more feasible, but leads to a number of issues such as collision-induced relaxation resulting in trap loss, as discussed in Section~\ref{sec:trapping}.

Molecules with partially filled electronic shells exhibit the Zeeman effect. A magnetic field interacts with the electronic spin and orbital angular momenta, which, if coupled to the molecular axis,  results in molecular alignment. Since the magnetic-field operator couples states of the same parity, molecules in a magnetic field are aligned, but not oriented. Semiclassically this can be thought of as a linear combination of magnetic moment projections in both directions along the molecular axis.

The Zeeman interaction operator can be generally written as~\cite{TownesSchawlow}
\begin{equation}
\label{ZeemanInt}
	V_m =  (g_L \Lambda + g_S \Sigma)\frac{M \Omega \mu_B B}{J(J+1)},
\end{equation}
where $g_L = 1$ and $g_S \approx 2$ are the orbital and spin gyromagnetic ratios, $\mu_B$ is the Bohr magneton, $B$ is the magnetic field magnitude, $M$ is the projection of the total angular momentum $\mathbf{J}$ on the laboratory $Z$-axis,  $\Lambda$ and $\Sigma$ are the projections of the electron orbital and spin angular momenta onto the molecular axis, and $\Omega = \Lambda+\Sigma$.
While a magnetic field hybridizes states of the same parity, it may bring states of opposite parity close together. When this happens, the dynamical properties of molecules, such as molecular orientation~\cite{BocaFri00, FriHerPCCP00} and collision dynamics \cite{KremsPRA03, KremsPRA03b, KremsDalgarnoJCP04, TscherbulPRL06, AbrahamssonJCP07,  TscherbulNJP09, CampbellDoylePRL09, TscherbulKremsFD09, SuleimanovJCP12, LemFriPRA09}, become very sensitive to external electric fields. 

The action of a far-detuned optical field on molecular rotational states was first considered by Zon and Katsnelson~\cite{Zon76}, and independently by Friedrich and Herschbach~\cite{FriHerPRL95, FriHerJPC95, FriHerZPhys96}.
In a far-off-resonant radiative field of intensity $I$, the rotational levels of a molecule undergo a dynamic Stark shift, as given by the Hamiltonian~\cite{CraigThiruBook}
\begin{equation}
	\label{Hi}
	H_{} = B \mathbf{J}^2 - g(t) \frac{I}{2 c \varepsilon_0} e_j e_l^\ast \alpha^{\text{lab}}_{jl}(\omega),
\end{equation}
where  $e_j$ and $ e_l$ are the polarizations of incoming and outgoing photons, $\alpha^{\text{lab}}_{jl}(\omega)$ is the dynamic (frequency-dependent) polarizability in the laboratory frame, and $g(t)$ gives the time-profile of the pulse. We note that higher-order terms in the multipole expansion of the potential, such as second-order hyperpolarizability, pertaining to the 4th power of the field strength, are likely negligible at laser intensities below $10^{12}$ W/cm$^2$, see, e.g., Ref.~\cite{RenardPRL03}.

For a linear molecule the only nonzero polarizability components in the molecular frame are $\alpha_{zz}=\alpha_\parallel$ and $\alpha_{xx}=\alpha_{yy}=\alpha_\perp$. In the case of a linearly polarized laser field, Hamiltonian~(\ref{Hi}) can be recast as (in units of $B_e$):
\begin{equation}
	\label{Halpha}
	H_{} =  \mathbf{J}^2  - g(t) \Delta \eta (\omega) \cos^2 \theta_{} - g(t) \eta_\perp (\omega),
\end{equation}
where $\theta$ is the angle between the molecular axis and the polarization vector of the laser field. Thus, because of the azimuthal symmetry about the field vector, the induced dipole potential involves just the polar angle $\theta$ between the molecular axis and the polarization axis of the laser pulse. The dimensionless interaction parameter $\Delta \eta(\omega) =  \eta_\parallel(\omega) - \eta_\perp (\omega)$ with  $\eta_{\parallel,\perp} (\omega) =  \alpha_{\parallel, \perp} (\omega) I /(2\varepsilon_0 c B_e)$. We note that Eq.~(\ref{Halpha}) was derived in Refs.~\cite{FriHerPRL95, FriHerJPC95} using the semiclassical approach and the rotating wave approximation. 

The polarization vector of an optical field defines an axis of cylindrical symmetry, $Z$. The projection, $M$, of the angular momentum $\mathbf{J}$ on $Z$ is then a good quantum number, while $J$ is not. However, one can again use the value of $J$ of the field-free rotational state that adiabatically correlates with the hybrid state as a label, denoted by $\tilde{J}$, so that $|\tilde{J}, M; \Delta \eta \rangle \to |{J, M}\rangle$ for $\Delta \eta \to 0$. For clarity, we also label the values of $\tilde{J}$ by the tilde, e.g.\ with $\tilde{0}$ corresponding to $\tilde{J}=0$.

The induced-dipole interaction~(\ref{Halpha}) preserves parity, hybridizing states with even or odd values of $J$,
\begin{equation}
	\label{PendularState}
	|\tilde{J}, M; \Delta \eta \rangle = \sum_{J} c_{J}^{\tilde{J} M} (\Delta \eta ) |{J M}\rangle , \hspace{0.2cm} J+\tilde{J} \hspace{0.15cm} \text{even},
\end{equation}
and therefore aligns molecules in the laboratory frame. Aligned molecules do not possess a space-fixed dipole moment, in contrast to species oriented by an electrostatic field. We note that at the far-off-resonant wavelengths ($\sim$1000 nm) usually employed in alignment and trapping experiments, the dynamic polarizability $\alpha_{ij} (k)$ approaches its static limit, $\alpha_{ij} (0)$, for a number of molecules, e.g.\ CO, N$_2$, and OCS. However, this is not the case for alkali dimers having low-lying excited $^1\Sigma$ and $^1\Pi$ states, such as KRb and RbCs. Virtual transitions to these states contribute to the ground-state dynamic polarizability, rendering it a few times larger than the static value~\cite{KotochigovaPRA10, DeiglmayrDulieuJCP08}.

A far-off-resonant optical field of sufficiently large intensity leads to the formation of ``tunneling doublets" -- close-lying states of opposite parity with the same $M$ and $\Delta \tilde{J} = 1$~\cite{FriHerPRL95}. The doublet states can be mixed by extremely weak electrostatic fields, which paves the way to achieving strong molecular orientation in the laboratory frame~\cite{FriHerJCP99, FriHerJPCA99}. The energy gap between neighboring doublets increases with the field intensity, and is proportional to $2\sqrt{\Delta\eta}$ in the strong-field limit~\cite{HaerteltFriedrichJCP08}. 
As we demonstrate below, at large $\Delta\eta$ interaction between two ground-state molecules can be described within the lowest tunneling doublet~\cite{LemeshkoPRA11Optical}. Detailed theory of combined action of electrostatic and  laser fields with noncollinear polarizations has been elaborated in refs.~\cite{HaerteltFriedrichJCP08, OmisteJCP11, OmistePCCP11, OmistePRA12, NielsenPRL12l} for continuous-wave and pulsed laser fields.



As discussed in Section~\ref{sec:nonadiabatic}, molecules can also be aligned by short laser pulses. In the nonadiabatic regime a laser pulse leads to the formation of a rotational wave packet with time-dependent coefficients,
\begin{equation}
\psi (\Delta \eta (t))=\sum_{J}c_{J}(\Delta \eta  (t))|J,M\rangle \exp
\left( -\frac{iE_{J}t}{\hbar }\right)
\end{equation}
where $E_{J}$ labels the energies of the field-free $J$-states. The alignment cosine, $\langle \cos^2 \theta \rangle (t)$, also becomes time-dependent and the molecules exhibit ``revivals'' -- nonzero alignment after a pulse has passed.
The effect of combined electrostatic field and short laser pulses was first studied by Cai~\textit{et al.} \cite{CaiFriedrichPRL01}, and subsequently by a number of theorists and experimentalists, see Sec.~\ref{sec:align}.

\subsection{Long-range intermolecular interactions}
\label{Molec:interactions}

Interactions between molecules at large intermolecular separations are particularly important for the two-body, few-body and many-body dynamics at low temperatures. In fact, the work of many authors including Bohn and coworkers \cite{AvdeenkovBohnPRL03, AvdeenkovPRA04, TicknorBohnPRA05, AvdeenkovPRA06, BohnArxiv13}, Micheli and coworkers \cite{BuchlerZollerPRL07, MicheliPRA07}, Gorshkov and coworkers \cite{GorshkovPRL08b}, and Lemeshko and Friedrich~\cite{LemeshkoPRA11Optical, LemFri11OpticalLong} aiming at the suppression of inelastic scattering at ultralow temperatures relies on the controllability of the long-range dipole-dipole interactions, which play the dominant role in the interaction potential between polar molecules separated by a large distance~\cite{MirandaYeJin11}. For non-polar molecules, the interaction potential at large intermolecular distances is usually dominated by the quadrupole-quadrupole interactions, which can also be tuned by external fields~\cite{byrdJCP11, ByrdPRL12,  ByrdPRA12, BhongalePRL13}. A gas of molecules with tunable long-range interactions can be exploited to study novel many-body physics, as discussed in Section~\ref{sec:many-body}. 

The analytical form of the long-range intermolecular interactions can be obtained using the multipole expansion of the electrostatic interaction between the electrons and nuclei of two molecules separated by a distance $r$, generally written as \cite{ChangRMP67, StoneBook13, JeziorskiCR94, ChalasinskiCR94}

\begin{equation}
	\label{MultipoleExpansion}
	V = \sum_{n_1,n_2=0}^{\infty} \sum_{\nu \lambda} (-1)^{\nu+\lambda}  \frac{A_n (n_1, n_2)}{r^{n+1}} C(n_1 n_2 n; \nu \lambda~\nu+\lambda) Y_{n_1 \nu}\left(\theta_1,\phi_1\right) Y_{n_2 \lambda} \left(\theta_2,\phi_2\right)
	Y_{n, -\nu-\lambda}\left(\theta,\phi\right),
\end{equation}
where
\begin{equation}
	\label{AlphaCoeff}
	A_n (n_1, n_2)=(4\pi)^{\frac{3}{2}} \frac{(-1)^{n_2}}{(2n+1)} \left( \frac{(2n+1)!}{(2n_1+1)! (2n_2+1)!} \right )^{\frac{1}{2}} Q_{n_1} Q_{n_2},
\end{equation}
$C(J_1 J_2 J, M_1 M_2 M)$ are the Clebsch-Gordan coefficients~\cite{VarshalovichAngMom, ZareAngMom}, $Q_{n_i}$represents the $n_i$-th multipole moment of molecule $i$ ($i=1,2$), and $n=n_1+n_2$. The angles $(\theta_{1,2},\phi_{1,2})$ and $(\theta, \phi)$ give the orientation of the molecular axes and the intermolecular radius-vector in the laboratory frame, respectively. The lowest order non-zero multipole for a neutral  heteronuclear diatomic molecule is the dipole moment corresponding to $n_i=1$. For a homonuclear diatomic molecule, it is the quadrupole moment with $n_i = 2$. When $r$ is large, the interaction of Eq.~(\ref{MultipoleExpansion}) can be treated as a perturbation. The long-range interaction between neutral molecules  is thus determined by the matrix elements of the operator (\ref{MultipoleExpansion}), which can be evaluated with high accuracy if the eigenstates of the isolated molecules are known; in this article we do not discuss retardation effects.

Eq. (\ref{MultipoleExpansion}) and the rules of angular momentum algebra \cite{vanderAvoirdTCC80, HeijmenMolPhys96, VarshalovichAngMom, ZareAngMom} can be used to establish the following useful results. 
If molecules are in states with $J=0$, the diagonal long-range interaction is given by the van der Waals potential $\sim C_6/r^6$ (plus higher order terms). 
For molecules in states with $J > 1$, the dominant contribution to the long-range potential is determined by the quadrupole-quadrupole interaction, decaying as $\sim C_5/r^5$. 
In the limit of ultracold temperatures, the effect of intermolecular potentials with the long-range van der Waals or quadrupole-quadrupole interactions can be described by the $s$-wave scattering length $a$. The scattering length determines the effective contact  interaction $(4 \pi \hbar^2 a/m) \delta(\mathbf{r})$ which enters as a pseudopotential in the many-body Gross-Pitaevskii equation.  

If the molecules are prepared in superpositions of different parity states, the intermolecular potential at large $r$ is dominated by the dipole-dipole interaction. The dipole-dipole interaction decays as $\sim 1/r^3$ and possesses an anisotropic angular dependence. The contact interaction is no longer sufficient to parametrize the system and it is necessary to introduce an anisotropic pseudopotential \cite{DereviankoPRA03}. In the studies of molecular Bose-Einstein condensates, the pseudopotential is often written as a sum of the contact interaction, representing the short-range part of the interaction, and an explicit dipole-dipole interaction term derived from the multipole expansion \cite{RonenPRA06, ZillichPCCP11}. This is equivalent to the Born approximation and is accurate for most systems of interest.

Dipole-dipole interactions can be created (and tuned) by applying a dc electric field, which orients polar molecules in the laboratory frame, Eq.~(\ref{WFRigRot}). One can also engineer tunable dipole-dipole interactions by dressing molecules with microwave fields~\cite{LemeshkoPRL12}. In this case, a microwave field is applied to couple the rotational states of different parity (e.g.\ $J=0$ and $J=1$). While the time average of the laboratory frame dipole moment thus created is zero, the dipole moment is non zero in the rotating frame. Since all molecules oscillate in phase, this gives rise to non-zero dipole-dipole interactions. This effect can be used to tune the magnitude and sign of the dipole-dipole interactions   \cite{BuchlerZollerPRL07, MicheliPRA07, GorshkovPRL08b} and achieve the dipole blockade of microwave excitations in ensembles of polar molecules~\cite{LemeshkoPRL12}.

It is worth noting that the term ``long-range'', although used in physical chemistry for any interactions at large $r$, needs to be redefined in the context of ultracold scattering and many-body physics of ultracold molecules.  The ``long-range'' potentials $V(\mathbf{r})$ correspond to a diverging integral $\int V(\mathbf{r}) d \mathbf{r}$. In this sense, the dipole-dipole interaction is long-range in three dimensions, but short range in two and one dimensions. In the absence of screening~\cite{LiebRMP76}, in systems with long range interactions the energy per particle depends not only on the particle density, but also on the total number of particles.

\vspace{30.pt}


\section{Field control of molecular beams}
\label{sec:beams}

The molecular beam technique was pioneered in 1911 by Dunoyer~\cite{Dunoyer1911}, who demonstrated that sodium atoms travel in vacuum along straight lines and thereby confirmed one of the key assumptions of the kinetic theory of gases. Ten years later Kallmann and Reiche from the Kaiser Wilhelm Institute for Physical Chemistry and Electrochemistry in Berlin~\footnote{Now Fritz Haber Institute of the Max Planck Society}  proposed to deflect a beam of polar molecules using an inhomogeneous electric field. The goal of the experiment was to determine whether a dipole moment is a property of individual molecules or if it arises only in the bulk due to the intermolecular interactions~\cite{KallmannReiche21}. The article of Kallmann and Reiche prompted Stern to publish a proposal describing his ongoing experiment~\cite{Stern21}, which later became the celebrated Stern-Gerlach experiment  on spatial quantization, based on deflection of an atomic beam in an inhomogeneous magnetic field~\cite{GerlachSternZPhys22}.  The first experiment on electric field deflection of polar molecules was performed by Stern's graduate student Wrede several years later~\cite{WredeZPhys27}. 

In the following decades, control over the transverse molecular motion was achieved by developing focusing techniques for molecular beams. In 1939, Rabi proposed the molecular beam magnetic resonance method that allowed for accurate measurements of magnetic dipole moments~\cite{RabiPR39}. In the 1950's, Bennewitz and coworkers used electric fields to focus a molecular beam onto a detector~\cite{BennewitzZphys55}, and Townes and coworkers used an electric quadrupole to focus a  state-selected beam of ammonia molecules to a microwave cavity, a key ingredient  of the maser~\cite{GordonPR54, GordonPR55}. Further technological developments allowed for control of more complex species.   The electrostatic deflection technique allowed to separate different stereoisomers of complex molecules~\cite{FilsingerAndChem09}. Alternating gradient focusing technique provided a means to transversely confine high-field seeking molecules, including polyatomic species~\cite{AuerbachJCP66, KakatiPLA67,  KakatiPLA69, KakatiJPE71, GuntherZPhys72, LubbertCPL75, LubbertJCP78, BethlemPRL02, TarbuttPRL04, BethlemJPB06, FilsingerPRL08, WohlfartPRA08b, KupperFarDiss09, KalninsRSI02}, such as benzonitrile (C$_7$H$_5$N)~\cite{WohlfartPRA08}, and complexes formed between benzonitrile molecules and argon atoms~\cite{PutzkeJCP12}. The same technique was used for the development of electric field guides for polyatomic molecules in specific rotational states \cite{PutzkePCCP11}.  The alternating gradient technique also allowed selective control of structural isomers of neutral molecules~\cite{FilsingerPRL08,  FilsingerPRA10}, and spatial separation of state- and size-selected neutral clusters \cite{TrippelPRA12}. The velocity selection of molecules can be performed using a curved electrostatic guide~\cite{RangwalaPRA03, JunglenPRL04, RiegerPRL05, TsujiJPB10, BertschePRA10} or a rotating nozzle~\cite{GuptaJPCA01, McCarthyJCP06}. Strebel~\textit{et al.} demonstrated velocity selection of ammonia molecules with a rotating nozzle, followed by guiding using a charged wire~\cite{StrebelPRA11}. Magnetic deflection of molecular beams has been used for state analysis and selection, see Ref.~\cite{AquilantiIJMS95} and references therein.

When subject to an intense laser field, molecules acquire an induced dipole moment. This opens a way to manipulate the motion of any polarizable molecules, not just of polar ones, by modulating the intensity of an applied laser field. Stapelfeldt~\textit{et al.}~\cite{StapelfeldtPRL97} demonstrated the deflection of a molecular beam by a gradient of an intense laser field. 
Optical deflection of I$_2$ and CS$_2$ molecules was also demonstrated by Sakai~\textit{et al.}~\cite{SakaiPRA98}.   
 Purcell and Barker measured the effect of molecular alignment on the optical dipole force acting on molecules, and proposed to use molecular axis polarization to control the translational motion~\cite{PurcellPRA10}.  Zhao~\textit{et al.} demonstrated a cylindrical lens for molecules formed by a nanosecond laser pulse~\cite{ZhaoPRL00} and a laser-formed ``molecular prism'' that allowed them to separate a mixture of benzene and nitric oxide~\cite{ZhaoJCP03}. The prism was also used to separate benzene and carbon disulfide molecules~\cite{ChungJCP01}. Averbukh~\textit{et al.} proposed and experimentally demonstrated spatial separation of molecular isotopes using nonadiabatic excitation of vibrational wavepackets~\cite{AverbukhPRL96, LeibscherPRA01}. The effects of laser-induced pre-alignment on the Stern-Gerlach deflection of paramagnetic molecules by an inhomogeneous static magnetic field were studied in Ref. \cite{GershnabelJCP11}. It was theoretically shown that using strong femtosecond laser pulses one can efficiently control molecular deflection in inhomogeneous laser fields~\cite{GershnabelPRL10}. Gershnabel and Averbukh investigated deflection of rotating rigid rotor and symmetric top molecules in inhomogeneous optical and static electric fields and discussed possible applications for molecular optics~\cite{GershnabelPRA10, FlossPRA11, GershnabelJCP11b, GershnabelJCP11}. Two experiments investigated the effect of microwave fields on molecules in a molecular beam~\cite{OdashimaPRL10, EnomotoAPB12}, opening the prospect for manipulation of molecular beams with low-frequency ac fields.  Quantum state dependent deflection of OCS molecules and characterization of the resulting single-state molecular beam by impulsive alignment was recently demonstrated by Nielsen \textit{et al.}~\cite{NielsenPCCP11}.

Although experiments on controlling the transverse motion of molecular beams date back to the work of Stern and Gerlach, 
control of the longitudinal motion remained a challenge until very recently. Electric field deceleration of neutral molecules was first attempted by John King at the Massachusetts Institute of Technology to produce a slow beam of ammonia in order to obtain a maser with an ultranarrow linewidth.
 At the University of Chicago,  Lennard Wharton constructed an 11-m-long molecular beam machine for the acceleration of LiF molecules in high-field-seeking states from 0.2 to 2.0 eV, aiming to produce high-energy molecular beams for reactive scattering studies. Unfortunately, both of these experiments were unsuccessful and were discontinued~\cite{AuerbachJCP66, WolfgangSciAm68}. Only in 1999,  Meijer  and coworkers demonstrated that the longitudinal motion of molecules can be controlled by the `Stark decelerator' -- they slowed a beam of metastable CO molecules from 225 m/s down to 98 m/s~\cite{BethlemPRL99}. Soon thereafter, Gould and coworkers presented a proof-of-principle experiment demonstrating the slowing of a beam of Cs atoms~\cite{MaddiPRA99}.

 In a Stark decelerator, an array of electrodes creates an inhomogeneous electric field with periodic minima and maxima of the field strength. 
Polar molecules travelling in this electric field potential lose kinetic energy, when climbing the potential hills, and gain kinetic energy, when going down the hills. To prevent kinetic energy gain, the electric field is temporally modulated so that the molecules most often travel against the gradient of the electric field potential. To obtain full insight into the dynamics of this process, the group of Meijer studied the phase stability of the Stark decelerator~\cite{BethlemPRL00, BethlemPRA02, vandeMeerakkerPRA05, vandeMeerakkerPRA06, GubbelsPRA06}, the group of Friedrich developed analytic models of the acceleration and deceleration dynamics~\cite{FriedrichEPJD04, GubbelsPRA06}, and Sawyer~\textit{et al.} provided a detailed study of the Stark deceleration efficiency \cite{SawyerEPJD08}.
After Stark deceleration was demonstrated for metastable CO ($a^3\Pi$) molecules~\cite{BethlemPRL99}, it was applied to a number of other molecules, including   $^{14,15}$ND$_3$~\cite{BethlemNature00, VeldhovenEPJD04}, NH$_3$~\cite{BethlemPRA02}, OH~\cite{BochinskiPRL03, BochinskiPRA04, vandeMeerakkerPRL05, ScharfenbergPRA09, KirsteJCP12}, OD~\cite{HoekstraPRL07}, NH ($a^1\Delta$)~\cite{MeerakkerJPB06, RiedelEPJD11}, NO~\cite{WangArxiv13}, H$_2$CO~\cite{HudsonPRA06}, SO$_2$~\cite{JungPRA06}, LiH~\cite{TokunagaNJP09}, CaF~\cite{WallPRA10}, YbF~\cite{BulleidPRA12}, and SrF~\cite{HoekstraPrivate, BergEPJD12}. 
Stark deceleration of 
 CH$_3$F~\cite{DengChinPB09} and CH~\cite{DengChinPLett09} has been discussed, though not yet realized. Translationally cold ground-state CO molecules were produced by optical pumping of Stark-decelerated metastable CO~\cite{BloklandJCP11}.

Detailed understanding of the deceleration dynamics allowed the development of methods for controlling the longitudinal motion of molecular beams on much smaller length scales. Marian~\textit{et al.}~\cite{MarianEPJD10} demonstrated an 11 cm long wire Stark decelerator, whereas Meek~\textit{et al.}~\cite{MeekPRL08, MeekScience09, MeekNJP09, MeekPRA11} developed a molecular chip decelerator -- a structure about 50 mm long -- and brought molecules to standstill. Later, molecular spectroscopy experiments were performed on a chip~\cite{SantambrogioCPC11, AbelMolPhys12}. Based on a similar idea, the microstructured reflection~\cite{SchulzPRL04} and focusing~\cite{GonzalezFlorezPCCP11} techniques were demonstrated. As a larger-scale analogue of the chip, deceleration of molecules in a macroscopic traveling potential was developed by Osterwalder and coworkers~\cite{OsterwalderPRA10, MeekRSI11, BulleidPRA12}. Quintero-P{\'e}rez \textit{et al.}~\cite{QuinteroPerez13} used the traveling-wave decelerator to bring the ammonia molecules to a standstill.

In many cases, Stark deceleration experiments are state-selective and decelerate molecules in low-field-seeking states.
 Extending the technique to molecules in high-field-seeking states has proven difficult, primarily because high-field-seekers are attracted to the electrodes, which mars the transverse stability of the beam. To overcome this problem, it has been proposed to use alternating gradient focusers~\cite{AuerbachJCP66, BethlemJPB06, KalninsRSI02}, leading to the demonstration of deceleration of metastable CO~\cite{BethlemPRL02}, YbF~\cite{TarbuttPRL04}, OH~\cite{WohlfartPRA08b}, and benzonitrile (C$_7$H$_5$N)~\cite{WohlfartPRA08} in high-field-seeking states, and guiding of low- and high-field-seeking states of ammonia~\cite{JunglenPRL04}. This is particularly important for polyatomic molecules, which in many cases cease to have low-field-seeking rotational states, due to a large density of states interacting with the field.

The Stark deceleration method, as demonstrated by Meijer and coworkers, is applicable only to polar molecules. 
To make the technique more general, Merkt, Softley, and their coworkers demonstrated an alternative method based on exciting molecules to a Rydberg state, which responds to an electric field due to an enormous dipole moment produced by the Rydberg electron and the ionic core. It was shown that a beam of H$_2$ molecules can be effectively slowed~\cite{YamakitaJCP04, SeilerPCCP11}.

Beams of paramagnetic atoms and molecules can be slowed using a Zeeman decelerator, which was applied to deceleration of neutral H and D atoms~\cite{VanhaeckePRA07, HoganPRA07, HoganJPB08}, metastable Ne atoms~\cite{NareviciusNJP07, NareviciusPRL08}, oxygen molecules~\cite{NareviciusPRA08, WiederkehrMolPhys12}, and methyl radicals~\cite{MomosePCCP13}. Phase stability in a Zeeman decelerator was studied by Wiederkehr~\textit{et al.}~\cite{WiederkehrPRA10}. Another type of a Zeeman decelerator recently demonstrated  is based on a moving, three-dimensional magnetic trap with tunable velocity~\cite{TrimecheEPJD11}. Narevicius~\textit{et al.}~\cite{NareviciusNJP07b} proposed to build a Zeeman decelerator using a series of quadrupole traps. This proposal was experimentally realized by Lavert-Ofir~\textit{et al.}~\cite{LavertOfirNJP11, LavertOfirPCCP11} for metastable Ne atoms.

Using the analogy with optical deflection of molecular beams, Friedrich proposed to use a gradient of an optical dipole force to slow molecules -- a so-called ``optical scoop''~\cite{FriedrichPRA00}. While the optical scoop technique has not yet been demonstrated experimentally, it was shown that molecules can be trapped in a moving periodic potential of a laser beam~\cite{MaherNatPhot12, FultonNatPhys06}. This technique allows one to trap, accelerate, and decelerate molecules preserving their narrow velocity spread, which amounts to an optical decelerator for molecules~\cite{FultonPRL04, BarkerPRA02}. Exploiting the ac Stark effect, Enomoto and Momose proposed a microwave decelerator~\cite{EnomotoPRA05}, which was experimentally realized by Merz~\textit{et al.}~\cite{MerzPRA12}.  Finally, Ahmad~\textit{et al.}~\cite{AhmadArxiv12} recently proposed to use coherent pulse trains for slowing multilevel atoms and molecules.

\vspace{0.5cm}

\section{External field traps for cold molecules}
\label{sec:trapping}

``If one extends the rules of two-dimensional focusing to three dimensions, one possesses all ingredients for particle trapping.'' While referring (mainly) to trapping of charged particles, these words of Wolfgang Paul in his Nobel address~\cite{PaulRMP90} also proved prophetic for cold molecules. One of the motivations behind the work on controlling the longitudinal motion of molecular beams was the prospect of slowing molecules to a standstill in order to confine them in a trap.   

  The external field traps exploit the dc Zeeman, dc Stark, or ac Stark effect in order to confine molecules in a particular quantum state by gradients of an applied field.  As a consequence of Maxwell's equations, it is possible to generate a dc field configuration, whether magnetic or electric, with a three-dimensional minimum, but not with a three-dimensional maximum. As a result,  dc magnetic and electric fields can only be used to confine neutral molecules in low-field-seeking states \cite{WingPRL80}. On the other hand, ac fields can be focused with a gradient of intensity leading to a three-dimensional maximum, which, depending on the detuning from resonance, can be used to confine molecules in either low-field-seeking states or high-field-seeking states. By preselecting molecules in a particular quantum state, external field traps effectively orient them, which can be used for a variety of applications such as controlled chemistry \cite{KremsPCCP08} or spectroscopy of large oriented molecules~\cite{DongSci02, KanyaCPL03, KanyaJCP04}. Confining molecules in an external field trap can provide long interrogation times for precision spectroscopy experiments compared to molecular beams~\cite{LevPRA06, SchnellFarDiss11, VeldhovenEPJD04, HudsonPRL06, BethlemEPST08, DemillePRL08b} and accurate measurements of the radiative lifetimes of vibrationally and electronically excited states~\cite{vandeMeerakkerPRL05b, CampbellPRL08, GilijamseJCP07}.  The latter is especially important for free radicals, for which there are few alternative methods.  Confining molecules in an external field is also necessary for sympathetic or evaporative cooling, currently the only methods for reaching quantum degeneracy with atomic ensembles. Ultracold molecules may open a new frontier in precision measurements leading to far-reaching applications. It is expected that experiments with ultracold molecules will provide a new, much improved, limit on the value of the electron electric dipole moment~\cite{DzubaIJMPE12, HudsonNature11, VuthaJPB10, VuthaJPB10} and the time variation, if any, of the fundamental constants~\cite{HudsonPRL06, DemillePRL08, ZelevinskyPRL08, BethlemFarDiss09, deNijsPRA12}.

Following the original proposal for buffer-gas loading of atoms and molecules into a magnetic trap~\cite{DoylePRA95, FriedrichFarDiss98}, Weinstein and coworkers reported the trapping of a neutral molecular radical CaH in a magnetic trap~\cite{WeinsteinCaH}. Beyond stimulating the research on cold molecules, magnetic trapping of molecules in the environment of an inert buffer gas became a useful tool for Zeeman spectroscopy~\cite{WeinsteinJCP98, PosthumusJPB99}, accurate measurement of the radiative lifetimes of long-lived molecular levels \cite{GilijamseJCP07} and the study of atomic and molecular collisions at temperatures near and below 1 Kelvin, see Sec.~\ref{sec:MolInteractions}. In addition to CaH, 
the experiments demonstrated the trapping of  CrH  and MnH~\cite{StollPRA08}  as well as of all the four stable isotopomers of NH radicals~\cite{CampbellPRL07, CampbellDoylePRL09, TsikataNJP10}. 
Long lifetimes of magnetically trapped molecules exceeding 20 seconds have been achieved~\cite{TsikataNJP10}. Although not widely used for molecules, trapping of magnetic species in the ground state is possible using an ac magnetic trap developed by Cornell~\textit{et al.}~\cite{CornellPRL91}.

  After achieving magnetic trapping, the number of various traps designed for molecules has quickly multiplied. 
Following the proposal by Wing~\cite{WingPRL80}, Jongma~\textit{et al.}~\cite{JongmaCPL97} proposed a scheme for trapping of CO molecules in an electrostatic trap. First electrostatic trapping experiment has been performed by Bethlem~\textit{et al.} for Stark-decelerated ammonia molecules~\cite{BethlemNature00}. In the follow-up studies, trapping of OH~\cite{vandeMeerakkerPRL05}, OD~\cite{HoekstraPRL07}, metastable CO~\cite{GilijamseJCP07}, and metastable NH~\cite{HoekstraPRA07}, has been demonstrated. Sawyer~\textit{et al.} demonstrated magneto-electrostatic~\cite{SawyerPRL07} and then permanent magnetic~\cite{SawyerPRL08} trapping of OH. Vanhaecke~\textit{et al.}~\cite{VanhaeckePRL02} demonstrated trapping of Cs$_2$ molecules in a quadrupole magnetic trap. 
Hogan~\textit{et al.}~\cite{HoganPRL08} demonstrated loading of magnetic atoms into a trap after Zeeman deceleration, a technique that can be likewise used for molecules. Kleinert \textit{et al.} \cite{KleinertPRL07} trapped NaCs molecules in a trap consisting of four thin wires creating a quadrupolar trapping field. The thin wires forming the electrodes of the trap allowed them to superimpose the trap onto a magneto-optical trap. Rieger \textit{et al.}~\cite{RiegerPRL05} reported an electrostatic trap consisting of five ring-shaped electrodes and two spherical electrodes at both ends. The trap was continuously loaded with a quadrupole guide, resulting in a steady-state population of trapped molecules. Kirste \textit{et al.}~\cite{KirstePRA09} trapped ND$_3$ molecules in the electrostatic analogue of a Ioffe-Pritchard trap. Van Veldhoven~\textit{et al.} demonstrated a cylindrical AC trap capable of trapping molecules in both low-field-seeking and high-field-seeking states~\cite{vanVeldhovenPRL05, BethlemPRA06}. Schnell~\textit{et al.} developed a linear AC trap for ground-state polar molecules~\cite{SchnellJPCA07, LutzowPRA08}.
A storage ring for neutral molecules was proposed by Katz~\cite{KatzJCP97} and later realized by Crompvoets~\textit{et al.}~\cite{CrompvoetsNature01}. The storage ring led to the development of the molecular synchrotron~\cite{HeinerNature07, HeinerPRA08, ZiegerPRL10}, i.e. a ring trap, in which neutral molecules can be accelerated or decelerated with high degree of control \cite{KremsNV07}.  An alternative design for a molecular synchrotron was proposed by Nishimura~\textit{et al.} \cite{NishimuraRSI03}. A microstructure trap for polar molecules was designed by Englert~\textit{et al.}~\cite{EnglertPRL11}. Confinement of OH radicals in a magnetoelectrostatic trap~\cite{SawyerPRL07}, which consists of a pair of coils located on the beam axis combined with an electrostatic quadrupole, as well as magnetic reflection by an array of magnets~\cite{MetsalaNJP08} was demonstrated with a Stark-decelerated beam. 
Accumulation of Stark-decelerated NH molecules in a magnetic trap was demonstrated by Riedel~\textit{et al.}~\cite{RiedelEPJD11}.

\begin{table}
{\footnotesize
\caption{Summary of the external field traps developed for neutral molecules to date. Only selected representative references are given. See text for a more comprehensive list of references.}
\begin{tabular}{| c | c | c | c | c | c| }
  \hline
  Trap type & molecule & trap depth (K) & molecule number & density (cm$^{-3}$) & Representative references \\
  \hline
    Electrostatic & OH, OD, CO $(a^3\Pi_1)$, NH$_3$, ND$_3$ & $0.1-1.2$ & $10^4-10^8$  & $10^6-10^8$ & \cite{vandeMeerakkerPRL05, HoekstraPRL07, RiegerPRL05, GilijamseJCP07, KirstePRA09, QuinteroPerez13} \\
  &  NH$(a^1\Delta)$, CH$_3$F, CH$_2$O, CH$_3$Cl  & & & & \cite{HoekstraPRA07, EnglertPRL11, ZeppenfeldNature12} \\
    Chip electrostatic & CO $(a^3\Pi_1)$ & $0.03-0.07$ & $10^3 - 10^4$  & $10^7$ ($10$/site) & \cite{MeekScience09} \\
  Thin-wire electrostatic & NaCs & $5\cdot 10^{-6}$ & $10^4-10^5$   & $10^5 - 10^6$ & \cite{KleinertPRL07} \\
    AC electric &$^{15}$ND$_3$ & $5\cdot10^{-3}$ & $10^3 - 10^4$  & $\sim 10^5$  & \cite{vanVeldhovenPRL05, BethlemPRA06, SchnellJPCA07, LutzowPRA08} \\
Magneto-electrostatic & OH & $0.4$ &  $10^5$   & $3\cdot10^3$ & \cite{SawyerPRL07} \\
Magnetic & OH & 0.2 &  $10^5$  & $10^6$ & \cite{SawyerPRL08, SawyerPCCP11} \\
Magnetic & CaH, NH & $\sim 1$ & $10^8 - 10^{10}$   & $10^7 - 10^{10}$ & \cite{WeinsteinCaH, CampbellPRL07, CollinThesis12, RiedelEPJD11} \\
Magnetic & CrH, MnH & $1$ & $10^5$   & $10^6$ & \cite{StollPRA08} \\
Magnetic & Cs$_2$ & $4\cdot10^{-6}$ & $10^5$   & $<10^8$ & \cite{VanhaeckePRL02} \\
Magnetic & KRb & $0.5$ &  $\sim 20-30$  & $10^4$ & \cite{WangPRL04} \\
Optical dipole trap  & alkali dimers, Sr$_2$ & $10^{-6}$ & $10^5$ & $10^{12}$ &  \cite{TakekoshiPRL98, JochimPRL03, HerbigSci03, JochimSci03, WinklerPRL05, ZirbelPRL08, NiScience08, OspekausPRL10, NiJinYeNature2010, OspelkausScience10, DebatinPCCP11, StellmerPRL12} \\
Optical lattice  & KRb, Cs$_2$, Sr$_2$  & $10^{-6}$ & $10^4$  & $(0.1-1)$/site &  \cite{ChotiaPRL12, MirandaYeJin11, DanzlNature10, DebatinPCCP11, StellmerPRL12, ReinaudiPRL12} \\
    \hline
\end{tabular}
}
\end{table}

The possibility of confining polarizable diatomic molecules by a laser field was proposed by Friedrich and Herschbach~\cite{FriHerPRL95}, and first demonstrated for Cs$_2$ molecules by Takekoshi~\textit{et al.}~\cite{TakekoshiPRL98}.  In subsequent years, optical trapping has become a widely used tool for trapping such species as KRb, LiCs, RbCs, as well as a broad range of homonuclear diatomic molecules, such as Cs$_2$, Sr$_2$, Rb$_2$, K$_2$~\cite{FriedrichCPC09, CarrNJP09, DulieuRPP09, KreStwFrieColdMol, ChinRMP10, StellmerPRL12, ReinaudiPRL12}.
Two-dimensional magneto-optical trapping of YO molecules was recently demonstrated by Hummon~\textit{et al.}~\cite{HummonPRL13}.
When prepared at ultracold temperatures, molecules can be trapped in a periodic potential of a laser beam -- optical lattice -- similarly to what had been previously achieved with ultracold atoms~\cite{LewensteinAdvPhys07, DanzlNature10, ChotiaPRL12, StellmerPRL12, ReinaudiPRL12}. 
In order to aid  experiments aiming to create controlled ensembles of ultracold molecules trapped in optical lattices,
Kotochigova and Tiesinga~\cite{KotochigovaPRA06} studied the possibility of manipulating ultracold molecules in optical lattices with microwave fields and Aldegunde~\textit{et al.}~\cite{AldegundePRA09} presented a detailed study of microwave spectra of cold polar molecules in the presence of electric and magnetic fields.
By analogy with optical traps, DeMille~\textit{et al.}~\cite{DemilleEPJD04} proposed to confine molecules in a microwave cavity. 
Due to small energy separation of the rotational energy levels and the low frequency of microwave fields, microwave traps have the promise of providing strong confinement over a large volume. Microwave trapping of molecules has yet to be realized.  
Table I summarizes the current state of the art in the development of external field traps for neutral molecules.

While the present article focuses on neutral molecules, it is necessary to acknowledge the tremendous recent progress in cooling and trapping of molecular ions. Due to their charge, even complex molecular ions can be trapped and cooled to milliKelvin temperatures~\cite{MolhavePRA00, BlythePRL05, RothJPB05, OstendorfPRL06, TongPRA11, WillitschIRPC12}. Just like neutral alkali metal dimers prepared from ultracold atoms, molecular 
 ions can be prepared in a selected rovibrational state by optical pumping~\cite{SchneiderNature10, StaanumNatPhys10} or sympathetic cooling of previously state selected ions~\cite{TongPRL10}. Molecular ions can be trapped for a few hours, with the rotational state lifetimes (for apolar ions) exceeding 15 minutes~\cite{TongPRL10}. 
The trapping and the accumulation of ions in specific rovibrational states can be used for highly accurate  rotational~\cite{ShenPRA12}, hyperfine~\cite{BresselPRL12}, and photodissociation~\cite{WellersPCCP11} spectroscopy, and provides a new approach to measuring the electric dipole moment of  electron~\cite{LeanhardtMolSpec11}.



\section{Controlling molecular rotation}
\label{sec:rotation}

\subsection{Molecular orientation in electrostatic fields}

The development of techniques for orienting and aligning molecules was largely motivated by the prospects of controlling molecular reaction dynamics. In field-free collision experiments, molecules rotate freely with the direction of the internuclear axes distributed uniformly in three-dimensional space. Restricting molecular rotations has long been a challenge. In the early 1960's Toennies and coworkers studied scattering of state-selected TlF molecules with a wide range of atomic and molecular collision partners in the presence of a homogeneous electrostatic field~\cite{BennewitzProc61, ToenniesFarDiss62, BennewitzZPhys64, ToenniesZPhys65, ToenniesZPhys66}. Experiments with symmetric top molecules, which possess nearly degenerate states of opposite parity and can, therefore, be effectively oriented by a weak electrostatic field of a hexapole, were pioneered by the groups of Brooks and Bernstein in the mid-1960's~\cite{ParkerBernsteinARPC, ChoJPC91, BrooksIRPC95, BrooksCPL79}. They used the hexapole technique to study  the effect of the orientation of reagents on the outcome of a number of chemical reactions, showing, e.g., that the chemical reactivity of K with CH$_3$I~\cite{BrooksJonesJCP66} and Rb with CH$_3$I~\cite{BeulerBernsteinJACS66, ParkerBernsteinCPL82} is substantially greater if the atom collides with the I end of the molecule. On the other hand, the results were different for the reaction of  K with CF$_3$I~\cite{Brooks69}, where the reaction favors the process of K attacking the CF$_3$ end of the molecule. Parker, Stolte and coworkers investigated the steric effects in collisions of He with CH$_3$F ~\cite{AdvancesMolecReactionsBook87}  and Ca with CH$_3$F~\cite{JannsenParkerStolteJPC91} as well as the influence of an electrostatic field on the reactivity of NO with O$_3$~\cite{vandenEndeStolteCP80} and Ba with N$_2$O~\cite{ParkerStolteJPC87,StolteBeamMethods88}. The hexapole technique was used to probe the changes of the integral cross sections in rotationally inelastic collisions with oriented reagents. By selecting the initial states with a hexapole field, and then orienting the molecular axis by a static electric field, ter Meulen and coworkers observed that O- and H-ended collisions of OH$(X)$ with Ar lead to different yields for low and high $J'$ levels~\cite{Beek00, vanBeekMeulenPRL01}. Stolte and coworkers observed an oscillatory dependence of the steric asymmetry in collisions of NO$(X)$ with He and Ar~\cite{vanLeukenCPL96, deLange99, deLangeStolteJCP04,GijsbertsenStoltePhysScr05}. Hexapole state selection was also used by Stolte and coworkers to measure the differential cross sections for collisions of He and D$_2$ with NO in a single $\Lambda$-doublet level~\cite{GijsbertsenStolteJCP05, GijsbertsenStolteJCP06}. Kaesdorf~\textit{et al.} studied angular-resolved photoelectron spectra of CH$_3$I oriented by an electric hexapole field~\cite{KaesdorfPRL85}. Kasai~\textit{et al.} used a 2-meter long electric hexapole to produce state-selected molecular beams of high intensity~\cite{KasaiRSI93}.

While the interaction of a rigid rotor molecule with an electric field
was theoretically studied in 1970 by von Meyenn~\cite{vonMeyennZPhys1970}, orienting molecules other than symmetric tops with external fields has, for a long time, been considered impractical. It was believed that orienting a molecular dipole in the laboratory frame would require an extremely high field strength~\cite{BernsteinHerschbachLevine87}. As an example we refer the reader to the paper by Brooks~\cite{BrooksScience76} published in Science in 1976, which contains a sections entitled `Brute
force -- how not to orient molecules'. Brooks referred to attempts to orient molecular dipoles by strong fields as the ``brute force methods'', and estimated the field strength required to suppress rotation and orient HCl to be 12 MV/cm, which is experimentally unfeasible. The ``brute force'' techniques were juxtaposed with the orientation methods based on choosing proper molecules, such as symmetric tops.

Only in the 1990's it was understood that it is possible to orient any polar molecule upon rotational cooling to very low temperatures ($<10$~K). The pioneering experiments were performed by Loesch and Remscheid who investigated the steric effect in K+CH$_3$I collisions~\cite{LoeschRemscheidJCP90}, and by Friedrich and Herschbach, who oriented a diatomic molecule (ICl) in a $^1\Sigma$ electronic state for the first time~\cite{FriHerNature91, FriedrichZPhys91}.  The effect of orientation of ICl on inelastic collisions with Ar was studied in Ref. \cite{Friedrich1992}.
Later, Loesch and coworkers investigated the influence of the molecular orientation on the velocity and angular distributions of the products in K+CH$_3$Br~\cite{vanLeukenLoesch95}, K+ICl~\cite{LoeschMollerJPC93}, Li+HF$\to$LiF+H~\cite{LoeschJCP93}, and K+C$_6$H$_5$I~\cite{LoeschJPCA97} reactions (for a review see Ref.~\cite{LoeschARPC95}). Friedrich and coworkers explored the effect of an electrostatic field on Ar+ICl collisions and found that the field suppresses rotationally inelastic scattering~\cite{FriedrichRubahnPRL92}. In a subsequent study,  Friedrich considered the thermodynamic properties of molecular ensembles in electrostatic and radiative fields~\cite{FriedrichCCCC93, FriedrichIRPC96, FriedrichEPJD06}.

The spectroscopic signature of pendular states resulting from molecule-field interactions were investigated theoretically and experimentally in subsequent articles~\cite{RostPRL92, FriedrichFarTrans93}. Block~\textit{et al.}~\cite{BlockPRL92} showed that a strong electrostatic field leads to collapse of the infrared spectrum of the (HCN)$_3$ molecule. Pendular states of ICl in an electrostatic field were detected via linear optical dichroism measurements by Slenczka~\cite{SlenczkaPRL98}. Pendular spectroscopy allowed Slenczka, Friedrich and Herschbach~\cite{SlenczkaCPL94} to measure the dipole moment of an excited-state of ICl($B^3\Pi_0$).
 Orientation of pyrimidine was measured spectroscopically~\cite{FranksJCP99}. A review of linear dichroism spectroscopy of large biological molecules oriented and aligned inside helium nanodroplets is presented in Ref.~\cite{KongIRPC09}. De~Miranda~\textit{et al.} performed the first experiment on controlling molecular orientation in the ultracold temperature regime~\cite{MirandaYeJin11}.

In addition to stereodynamics, the techniques for orienting molecules with external fields can be used for determining molecular properties. For example, Gijsbertsen and coworkers  \cite{GijsbertsenPRL07} showed that a measurement of the ion distribution produced by photodissociation of molecules oriented in a strong dc electric field can be used to determine the direction (sign) of the molecular dipole moment. In the case of NO, it was found that the dipole moment corresponds to N$^-$O$^+$. Gonz{\'a}lez-F{\'e}rez and Schmelcher developed techniques employing an electrostatic field to manipulate molecular rovibrational states~\cite{GonzalezFerezPRA04, GonzalezFerezEPL05, GonzalezFerezPRA05, GonzalezFerezCP06, GonzalezFerezNJP09}, control molecular association and dissociation~\cite{GonzalezFerezPRA07, GonzalezFerezPCCP11}, as well as radiative transition rates~\cite{MaylePRA07}. Melezhik and Schmelcher presented a method for ultracold molecule formation in a waveguide~\cite{MelezhikNJP09}.

{In the most recent developments, Rydberg molecules, consisting of a ground-state atom bound to a highly-excited atom were  predicted theoretically~\cite{GreenePRL00, ChibisovJPB02, HamiltonJPB02} and observed experimentally~\cite{BendkowskyNatPhys09}. These molecules are spatially very extended and possess a substantial value of a dipole moment, even in the case of homonuclear species~\cite{LiScience11}. Such molecules thus represent an attractive target for manipulation by electric and magnetic fields~\cite{LesanovskyJPB06, KurzEPL12}. Mayle~\textit{et al.}~\cite{MaylePRA12} proposed a mechanism for electric field control of ultralong-range triatomic polar Rydberg molecules. Rittenhouse and Sadeghpour predicted the existence of long-range polyatomic Rydberg molecules, consisting of a Rydberg atom and a polar dimer and developed techniques to coherently manipulate them using a Raman scheme~\cite{RittenhousePRL10}.}

\vspace{1cm}

\subsection{Molecular alignment in magnetic and laser fields}
\label{sec:align}

Apart from orientation (a single-headed arrow $\uparrow$), molecules can also be aligned (a double-headed arrow $\updownarrow$) along a particular space-fixed axis, not favoring one direction over another. In order to achieve alignment, one needs to apply a field preserving the parity of states such as a magnetic or a far-off-resonant laser field. The possibility to align paramagnetic molecules and ions in a magnetic field was first discussed by Friedrich and Herschbach~\cite{FriedrichZPhys92}, and then experimentally demonstrated for an excited state of ICl by Slenczka~\textit{et al.}~\cite{SlenczkaPRL94}.  A magnetic field hybridizes states of the same parity, but it may result in crossings between some states of opposite parity. This effect can be used to achieve strong orientation of molecules in combined magnetic and electric fields. Spectroscopy of ICl molecules in parallel strong electric and magnetic fields was performed in refs. \cite{FriedrichCPL94, FriedrichCanJP94}. The properties of $^2\Sigma$ and $^3\Sigma$ molecules in parallel electric and magnetic fields were investigated later in refs.~\cite{FriHerPCCP00, BocaFri00, KremsPRA03, KremsPRA03b, KremsDalgarnoJCP04, TscherbulPRL06, AbrahamssonJCP07,  TscherbulNJP09, CampbellDoylePRL09,  TscherbulKremsFD09, SuleimanovJCP12, LemFriPRA09, VuthaPRA11, SteimlePRA11, BickmanPRA09,  ShumanPRL09, ShumanNature10, BarryPRL12}.  It was shown that ensembles of polar paramagnetic molecules in combined static fields can be used for accurate 2D mapping of an ac field within a broad range of frequencies~\cite{AlyabPRA12}.

Zare~\cite{ZareBunsen82} proposed an alternative method of creating aligned molecules by exciting them with linearly polarized light. 
This technique was used by Loesch and Stienkemeier~\cite{LoeschJPC94} to study the reactions Sr+HF$\to$SrF+H and  K+HF$\to$KF+H. It was found that the reactivity of Sr with HF at low energies is greater when the HF bond is perpendicular to the approach direction of Sr, whereas the reactivity of the K+HF system is greater when the HF bond is parallel to the approach direction. At higher collision energies, the reactivity of Sr with HF becomes more pronounced for the collinear geometry, whereas the reactivity of K with HF is insensitive to the orientation of the HF bond. Zare and coworkers studied the steric effect in reactions of Cl with vibrationally excited CH$_4$ and CHD$_3$ by varying the direction of approach of the Cl atom relative to the C--H stretching bond with different polarizations of the infrared excitation laser \cite{SimpsonJCP1995b}. 

Strong molecular polarization can also be achieved by removing the population of certain $M$ sublevels by selective photodissociation. This technique was pioneered by de~Vries~\textit{et al.}~\cite{deVriesMartin83} who created a polarized beam of IBr to study the reaction of IBr with metastable Xe$^\ast$ producing XeI$^\ast$ and XeBr$^\ast$. It was found that the reaction cross section is larger when the Xe$^\ast$ atom approaches parallel to the plane of rotation of IBr, and smaller for the perpendicular approach direction. Finally, a coherent superposition of parity states can be prepared by a two-color phase locked laser excitation~\cite{VrakkingCPL97}. Qi~\textit{et al.} showed that Autler-Townes effect  can be used to achieve all-optical molecular alignment~\cite{QiPRL99}.

With the advance of optical technology it has been realized that the rotational motion of molecules can be steered using an intense laser field. 
Scattering of intense far-off-resonant laser light from symmetric-top molecules was first studied by Zon and Katsnelson in the mid-70's~\cite{Zon76}. Some 20 years later, unaware of this work, Friedrich and Herschbach~\cite{FriHerPRL95} presented a theoretical description of alignment of buffer-gas-cooled molecules by a far-detuned laser field, based on the experimental evidence provided by Normand~\textit{et al.} \cite{NormandJPB92}.  The possibility of alignment, focusing, and trapping of molecules in intense laser fields was studied in detail in the subsequent work of Friedrich and Herschbach \cite{FriHerJPC95, FriHerZPhys96}, and Seideman~\cite{SeidemanJCP95, SeidemanJCP97, SeidemanJCP97b, SeidemanPRA97, SeidemanJCP99, ZonEPJD00}. Friedrich and Herschbach studied the spectroscopic signatures of pendular states and collapse of infrared spectrum~\cite{FriedrichCPL96}.  Dion~\textit{et al.}~\cite{DionPRA99} performed a numerical study of the effects of permanent and induced dipole moments on alignment. 

The theoretical proposals prompted the development of novel experiments on molecular alignment. Kim and Felker spectroscopically detected the pendular states resulting from alignment of the naphthalene trimer molecules and benzene-argon complexes~\cite{KimJCP96, KimJCP97, KimJCP98}. Mathur and coworkers detected the pendular states by measuring ion distributions after photodissociation of CS$_2$, CO$_2$, NO$_2$, H$_2$O, CCl$_4$, CHCl$_3$, and CH$_2$Cl$_2$  molecules by an intense linearly polarized picosecond pulse~\cite{KumarJPB96, KumarPRA96, SafvanJPB99, BhardwajPRA97, BhardwajJPB99}. Posthumus~\textit{et al.}~\cite{PosthumusJPB99} performed a double-pulse measurement of molecular alignment. The first direct evidence of adiabatic molecular alignment and its quantitative characterization by measuring photofragment distributions were reported in the seminal papers of Sakai, Stapelfeldt, and coworkers, for I$_2$, ICl, CS$_2$, CH$_3$I, and C$_6$H$_5$I molecules~\cite{SakaiJCP99, LarsenJCP99}, also see Ref.~\cite{SugitaJCP00}.  Ji~\textit{et al.} studied final state alignment in all-optical multiple resonance quantum-state selected photodissociation of K$_2$~\cite{JiPRA94, JiJCP95}.
Larsen \textit{et al.}~\cite{LarsenPRL99} performed controlled photodissociation of  I$_2$ molecules, manipulating the branching ratio of the I$+$I and I$^\ast +$I product channels by aligning the molecular axis.   The role of rotational temperature in adiabatic molecular alignment was studied experimentally in Ref.~\cite{KumarappanJCP06}.  Pentlehner~\textit{et al.} studied alignment of C$_6$H$_4$I$_2$, C$_6$H$_5$I, and CH$_3$I molecules dissolved in helium nanodroplets~\cite{PentlehnerPRA13}.

In the case of a polyatomic molecule, alignment by a single linearly polarized laser field leads to 1D alignment, i.e.\ only one molecular axis becomes aligned. Achieving a complete 3D alignment had been a challenging task and was first demonstrated by Larsen \textit{et al.}~\cite{LarsenPRL00} for $3,4-$dibromothiophene molecules in an elliptically polarized laser field. The theory of 3D alignment and external field control of the torsional motion in polyatomic molecules was developed by Seideman and coworkers~\cite{RamakrishnaPRL07b, ArtamonovJCP08}. Manipulating torsion in molecules using laser pulses was demonstrated experimentally and theoretically in refs.~\cite{MadsenPRL09, MadsenJCP09, HansenJCP2012}. A method of coherent control of torsion in biphenyl derivative using laser fields was proposed in Ref.~\cite{ParkerJCP11}. Spence and Doak proposed to use alignment to perform structural studies of small proteins via single-molecule X-ray diffraction~\cite{SpencePRL04}.  State selection and manipulation of molecular beams for the applications of diffraction studies was discussed in Ref.~\cite{FilsingerPCCP11}.
 Ortigoso discussed the possibility of using microwave pulses to engineer coherent rotational states in asymmetric top molecules~\cite{OrtigosoPRA98}.

If the linear polarization of an intense laser is slowly rotating, the molecular alignment must adiabatically follow it; thus a laser field can act as an ``optical centrifuge'' for molecules. Speeding up the rotation makes it possible to prepare molecules in rotational states with extremely high angular momenta, all the way to the dissociation limit, as demonstrated for Cl$_2$ in Ref.~\cite{KarczmarekPRL99, VilleneuvePRL00}. The possibility to use optical centrifuge for selective dissociation was studied in subsequent papers~\cite{SpannerJCP01, SpannerJCP01b, HasbaniJCP02}.  The difference in molecular moments of inertia require different torques for dissociation, making it possible to separate molecules composed of different isotopes, such as $^{35}$Cl$_2$ and $^{37}$Cl$_2$. The optical centrifuge was used to study the dynamics and collisions of polyatomic molecules in high-energy rotational states~\cite{YuanFarDiss11, YuanPNAS11}. Motivated by the work of Forrey~\textit{et al.}~\cite{ForreyPRA01, ForreyPRA02, TilfordPRA04} Korobenko and coworkers~\cite{KorobenkoArxiv13} developed an optical centrifuge to study collisions of cold diatomic molecules in highly excited rotational states.
A cw laser approach for excitation of the highest possible rotational levels was also proposed for Li$_2$~\cite{LiJCP00}. 
 
The electron distribution in aligned molecules can be studied using strong-field ionization, as first demonstrated experimentally by Litvinyuk~\textit{et al.}~\cite{LitvinyukPRL03}. Dooley~\textit{et al.}~\cite{DooleyPRA03} performed direct imaging of rotational wave-packet dynamics  via ion imaging after Coulomb explosion  of N$_2$ and O$_2$ molecules.  The theory of alignment-dependent ionization probability of molecules in a double-pulse laser field was developed by Zhao \textit{et al.}~\cite{ZhaoPRA03}. Suzuki~\textit{et al.}~\cite{SuzukiPRL04} developed a technique to control multiphoton ionization of aligned I$_2$ molecules using time-dependent polarization pulses.
Intense laser ionization of transiently aligned CO was studied in Ref.~\cite{PinkhamPRA05}. Kumarappan~\textit{et al.}~\cite{KumarappanPRL08} measured the electron angular distributions resulting from multi-photon ionization of laser-aligned CS$_2$ molecules.  Ionization of one- and three-dimensionally-oriented benzonitrile~\cite{HansenPRA11} and oriented OCS~\cite{DimitrovskiPRA11} molecules by intense circularly polarized femtosecond laser pulses provided insight into the structure of molecular orbitals.

Recombination of electrons appearing due to strong-field ionization results in high-harmonic generation (HHG)~\cite{BrabecRMP00}. In aligned molecules, high-order harmonic generation leads to quantum interference~\cite{KanaiNature05, KatoPRA11, SakemiPRA12, ZhouPRA05, VozziPRL05}, which can be used to study the role of orbital symmetry~\cite{NaldaPRA04}, 
and other electronic and rotational properties of molecules ~\cite{RamakrishnaPRL07}.
Itatani~\textit{et al.}~\cite{ItataniNature04} used  HHG for imaging the molecular electron orbitals of N$_2$. Wagner~\textit{et al.}~\cite{WagnerPNAS06} studied the molecular dynamics using coherent electrons from HHG. The dependence of the HHG polarization and ellipticity on molecular alignment was elucidated by Levesque and coworkers~\cite{LevesquePRL07} and Kanai and coworkers~\cite{KanaiPRL07}. The possibility to observe ellipticity in higher order harmonic generation from aligned molecules and its relations to the underlying molecular potential were studied in refs.~\cite{RamakrishnaPRA10, SherrattPRA11}.  Kelkensberg~\textit{et al.}~\cite{KelkensbergPRA11} measured photoelectron angular distribution from aligned molecules interacting with intense XUV pulses, which revealed contributions from the different orbitals of a CO$_2$ molecule.  Lock~\textit{et al.}~\cite{LockPRL12} demonstrated experimentally and theoretically that high-harmonic generation can be used as a sensitive probe of the rotational wave packet dynamics. Bartels~\textit{et al.}~\cite{BartelsPRL01} showed that the time-dependent phase modulation induced by molecular rotational wave packets can be used to manipulate the phase and spectral content of ultrashort light pulses (e.g. increase the bandwidth of the pulse) \cite{BartelsPRL01}. Frumker and coworkers~\cite{FrumkerPRL12} developed a method to study the dynamics of oriented molecules by detecting harmonic radiation.


Laser-induced molecular alignment techniques can also be used for measuring polarizability anisotropies of rare gas diatomic molecules~\cite{MinemotoJCP03, MinemotoJCP11}, to control surface phenomena such as surface reactions~\cite{ShreenivasJPCA10}, and self-assembly of aligned molecules on liquid surfaces~\cite{NevoJCP09}. Applications of aligned and oriented molecules to study elementary processes on surfaces was discussed by Vattuone \textit{et al.}~\cite{VattuonePrSurfSci10}. It was shown that laser-induced torsional alignment of nonrigid molecules can be performed in a nuclear-spin-selective way, which allows for selective manipulation of nuclear spin isomers~\cite{FlossJCP12}. Kiljunen~\textit{et al.} studied the effect of a crystal field on alignment of matrix-isolated species~\cite{KiljunenPRL05, KiljunenPRA05, KiljunenJCP06}. The alignment of organic molecules adsorbed on a semiconductor surface was proposed to realize a fast molecular switch for conductance \cite{ReuterPRL08,ReuterPRA12}. 
Khodorkovsky~\textit{et al.}~\cite{KhodorkovskyPRA11} proposed techniques to modify the molecule-surface scattering using laser pulses. The possibility to use tailored laser pulses to control molecular machines was discussed in Ref.~\cite{PerezHernandezNJP10}. Artamonov and Seideman showed that simultaneous alignment and focusing of molecules can be achieved due to the field enhancement by surface plasmons of metal nanoparticles~\cite{ArtamonovNano10}.

 \vspace{0.5cm}
\subsection{Molecular orientation in combined fields}
\label{sec:combined}

Although a far-off-resonant laser field mixes only states of same parity, it leads to the formation of the so-called tunneling doublets -- closely lying states of opposite parity. In a way, these doublets are similar to tunneling doublets in ammonia or $\Lambda$-doublets in open-shell molecules, such as OH, and can be hybridized by very weak electrostatic fields leading to strong molecular orientation~\cite{TownesSchawlow}. The possibility to achieve enhanced molecular orientation in collinear laser and electrostatic fields was first theoretically considered by Friedrich and Herschbach~\cite{FriHerJCP99, FriHerJPCA99}. Later, H{\"a}rtelt and Friedrich provided a detailed study of symmetric-top states in electric and radiative fields with an arbitrary polarization tilt~\cite{HaerteltFriedrichJCP08}. The theory was extended to asymmetric tops by Omiste~\textit{et al.}~\cite{OmisteJCP11}. A theoretical description of laser alignment and combined-field orientation of benzonitrile was presented in Ref.~\cite{OmistePCCP11}. Sokolov~\textit{et al.} proposed a technique based on a combination of an infrared and ultraviolet laser pulses to achieve molecular orientation \cite{SokolovPRA09}.

Enhanced molecular orientation in combined electrostatic and laser fields was demonstrated by  Friedrich~\textit{et al.}~\cite{FriedrichJModOpt03, NahlerJCP03} for HXeI molecules, and by Sakai~\textit{et al.}~\cite{SakaiPRL03, MinemotoJCP03b} for OCS. Tanji~\textit{et al.}~\cite{TanjiPRA05} demonstrated the possibility of three-dimensional orientation of polyatomic molecules by combined electrostatic and elliptically polarized laser fields. Orientation of a large organoxenon (H--Xe--CCH)  molecule in combined laser and electrostatic fields was recently demonstrated in Ref. \cite{PoteryaJCP08}. Trippel~\textit{et al.}~\cite{TrippelMolPhys13} demonstrated strong alignment and orientation of molecular samples at kHz repetition rate.

In more exotic developments, it was shown that molecules in combined electrostatic and radiative fields can be described by supersymmetric quantum mechanics, which allows one to find analytically solvable cases of the combined-field problem~\cite{LemMusKaisFriPRA11, LemMusKaisFriLong}.

\vspace{0.5cm}
\subsection{Interaction with short laser pulses: nonadiabatic alignment and orientation}
\label{sec:nonadiabatic}

Laser pulses shorter than the rotational period perturb the rotational energy states of a molecule nonadiabatically. When a molecule is subjected to such a short laser pulse, a rotational wave packet is formed.  The phases of the rotational states in the rotational wave packet evolve in time even after the pulse is gone, which manifests itself in revivals of alignment.  The idea of using short laser pulses for time-dependent molecular alignment was proposed by Ortigoso~\textit{et al.} \cite{OrtigosoFriedrich99} and elaborated by other authors~\cite{ SeidemanPRL99, CaiFriedrichCCCC01, LemFriJPCA10}.  The effect of thermal averaging on the post-pulse alignment and the experimental limitations to observe it were studied by Machholm~\cite{MachholmJCP01}. A number of authors~\cite{KellerPRA00, GranucciJCP04, TorresPRA05, LemFriJPCA10} presented numerical calculations for molecular dynamics during and after short and long pulses which illustrated the relative effect of the adiabatic and nonadiabatic interactions. Rotational wavepacket dynamics in the presence of dissipation leading to decoherence of rotational wave packets was investigated in refs.~\cite{PelzerJCP07, PelzerJCP08}.

It is worth noting, that the first signatures of the post-pulse alignment were observed in birefringence of CS$_2$ vapor back in 1975~\cite{HeritagePRL75}, furthermore, nonadiabatic excitation of rotational wavepackets had been widely used for rotational coherence spectroscopy~\cite{OhlineLC94}. However, the first experiment on nonadiabatic alignment of molecules in beams was performed in 2001 by Rosca-Pruna and Vrakking \cite{RoscaPrunaPRL01, RoscaPrunaJCP02a, RoscaPrunaJCP02b, RoscaPrunaJPB01, SpringateJPB01}, who observed revivals of the alignment of I$_2$ and Br$_2$ molecules. Miyazaki~\textit{et al.}~\cite{MiyazakiPRL05} observed field-free alignment of molecules using high-order harmonic generation. Later, field-free alignment of deuterium by femtosecond pulses was demonstrated~\cite{LeeJPB06}.  Sussman~\textit{et al.}~\cite{SussmanPRA06} proposed to align molecules by rapidly truncating a long laser pulse.

Very soon after the experiments with linear molecules, nonadiabatic alignment of symmetric top molecules (methyliodide and tert-butyliodide) by short laser pulses was demonstrated in Ref.~\cite{HamiltonPRA05}.  Nonadiabatic alignment and rotational revivals of asymmetric top molecules was demonstrated for the case of iodobenzene and iodopentafluorobenzene~\cite{PeronnePRL03, PeronnePRA04, PoulsenJCP04}. Field-free one-dimensional alignment of an asymmetric top  molecule C$_2$H$_4$ was studied in refs.~\cite{RouzeePRA06, XuJASMS06}. Control of rotational wave-packet dynamics in asymmetric top molecules was demonstrated in Ref.~\cite{HolmegaardPRL07}. Pentlehner~\textit{et al.} studied nonadiabatic alignment of CH$_3$I molecules dissolved in helium nanodroplets~\cite{PentlehnerPRL13}.

Three-dimensional field-free  alignment of ethylene was theoretically investigated by Underwood~\textit{et al.}~\cite{UnderwoodPRL05}. Experimentally, field-free three-dimensional alignment of polyatomic molecules was demonstrated be Lee~\textit{et al.}~\cite{LeePRL06} for the case of SO$_2$. Viftrup~\textit{et al.}~\cite{ViftrupPRL07} demonstrated an interesting technique for controlling 3D rotations of polyatomic molecules. In this method a long linearly polarized (nanosecond) pulse strongly aligns the most polarizable axis of an asymmetric top molecule along its polarization axis while an orthogonally polarized, short (femtosecond) pulse puts the molecules into controlled rotation about the axis of alignment. A laser-field method to control rotation of asymmetric top molecules in 3D, based on combined long and short laser pulses, was described in Ref.~\cite{ViftrupPRA09}.

A number of more sophisticated techniques were developed to achieve better control over the rotational motion of molecules. 
For example, Bisgaard~\textit{et al.}~\cite{BisgaardPRL04, BisgaardPRA06} proposed a technique to enhance molecular alignment by using two consecutive laser pulses. Poulsen~\textit{et al.}~\cite{PoulsenPRA06} demonstrated enhancement of alignment using two temporally overlapping laser pulses. A method for enhanced orientation of linear dipolar molecules using a half-cycle pulse combined with a delayed laser pulse inducing molecular anti-alignment was discussed in refs.~\cite{GershnabelPRA06b, GershnabelPRA06}. Improved alignment by  shaped laser pulses was demonstrated in refs.~\cite{HornPRA06, PinkhamPRA07, HertzPRA07}. It was shown theoretically and experimentally that a pair of linearly polarized fs pulses can orient rotational angular momentum on a ps timescale~\cite{KitanoPRL09}. Leibscher~\textit{et al.}~\cite{LeibscherPRL03, LeibscherPRA04} showed that the degree of alignment achievable with a single pulse is limited, and proposed a technique to  improve it using trains of short laser pulses.

The possibility to achieve post-pulse molecular orientation in the presence of a weak electrostatic field and a short laser pulse was considered by Cai~\textit{et al.}~\cite{CaiFriedrichPRL01}. Field-free molecular orientation in combined fields was first demonstrated for the case of OCS in Ref.~\cite{GobanPRL08}, also see refs.~\cite{SugawaraPRA08, MuramatsuPRA09}. Holmegaard~\textit{et al.}~\cite{HolmegaardPRL09} demonstrated laser-induced alignment and orientation of iodobenzene molecules in low-lying quantum states, preselected by electric field deflection. 
Laser-induced 3D alignment and orientation of large state-selected molecules was demonstrated in refs.~\cite{NevoPCCP09, FilsingerJCP09l}. Nielsen~\textit{et al.}~\cite{NielsenPRL12l} provided a recipe for achieving the best molecular orientation using combined fields, with the underlying theory presented in Ref.~\cite{OmistePRA12}.

The first proposal to use half-cycle pulses in order to generate molecular orientation as opposed to alignment was formulated in Ref.~\cite{MatosAbiaguePRA03}. The possibility of spatial orientation by a half-cycle laser pulse and a series of  fs pulses  to manipulate the vibrational dynamics of OHF$^-$ was studied theoretically in Ref.~\cite{ElghobashiMeinhardtJCP09}. An alternative way to achieve orientation using laser fields was proposed and experimentally realized by the group of Sakai~\cite{KanaiJCP01, OdaPRL2010}. They used two-color laser fields interacting with molecular anisotropic hyperpolarizability and anisotropic polarizability (without permanent-dipole moment interactions involved). Different mechanisms of field-free molecular orientation induced by two-color lasers were theoretically studied in refs.~\cite{SpannerPRL12, YunPRA11}. We note that nonadiabatic rotational dynamics and field-free orientation can also be studied by rapidly switching an electrostatic field, as discussed by S{\'a}nchez-Moreno~\textit{et al.} \cite{SanchezMorenoPRA07}.

De~\textit{et al.}~\cite{DePRL09} reported the first experimental observation of nonadiabatic field-free orientation of a heteronuclear diatomic molecule (CO) induced by an intense two-color (800 and 400 nm) femtosecond laser field. Field-free molecular orientation by nonresonant and quasiresonant two-color laser pulses was studied in Ref.~\cite{TehiniPRA08}.
Gu{\'e}run \textit{et al.}~\cite{GuerinPRA08} showed that a laser pulse designed as an adiabatic ramp followed by a kick allows one to reach a perfect post-pulse molecular alignment, free of saturation.
Ghafur \textit{et al.}~\cite{GhafurNat09} demonstrated post-pulse alignment and orientation of hexapole state-selected NO molecules using an electrostatic field and fs laser pulses. Nonadiabatic molecular orientation by polarization-gated ultrashort laser pulses was theoretically studied by Chen~\textit{et al.}~\cite{ChenPRA10}. 
Daems~\textit{et al.}~\cite{DaemsPRL05} showed  that a combination of a half-cycle pulse and a short nonresonant laser pulse produces a strongly enhanced post-pulse orientation.
Baek~\textit{et al.}~\cite{BaekJCP11} studied the field-free dynamics of benzene aligned by two consecutive short laser pulses, while the interaction of rotationally excited molecules with short laser pulses was studied by Owschimikow~\textit{et al.}~\cite{OwschimikowPRA09, OwschimikowJCP10, OwschimikowPCCP11}.
Yun~\textit{et al.}~\cite{YunPRA12} analyzed the time evolution of the quantum phases of the rotational states contributing to the rotational wave packet which provided new insights into the post-pulse alignment dynamics.

It was also shown that THz pulses are capable of exciting rotational wavepackets that exhibit revivals, which opens up a prospect for rotational spectroscopy~\cite{HardePRL91, BigourdOptComm08, FleischerPRL12} and the studies of molecular orientation~\cite{MachholmPRL01, ShuJCP10, QinPRA12}. 
Kitano~\textit{et al.} ~\cite{KitanoPRA11} proposed to apply an intense fs pulse creating a rotational wave packet and then a delayed  THz pulse that resonantly couples opposite parity states resulting in field-free orientation. Yu~\textit{et al.}~\cite{YuCP12} proposed to use two modulated few-cycle THz pulses to achieve long lived post-pulse molecular orientation. 

Since laser pulses can be tailored and modulated in many ways, one can design feedback control loops in order to prepare molecules in states of maximal alignment or orientation. For example, Suzuki~\textit{et al.}~\cite{SuzukiPRL08} demonstrated an optimal control technique for nonadiabatic alignment of  N$_2$ using shaped fs pulses with the feedback of the degree of alignment measured via ion imaging. With the development of high-intensity XUV sources, such as free-electron lasers, it became possible to probe the field-free molecular alignment via ionization and dissociation of molecules, mapping the fragments distribution~\cite{JohnssonJPB09}. Optimal control approach to the field free 3D alignment of an arbitrary (asymmetric top) molecule was developed in Ref.~\cite{ArtamonovPRA10}. Nakajima~\textit{et al.} developed an optimal control theory for the field-free molecular orientation by phase-locked two-color laser pulses~\cite{NakajimaJPCA12}.

Control over rotational motion opens the possibility to study molecular dynamics in new, interesting regimes. 
For example,  Fleischer~\textit{et al.}~\cite{FleischerNJP09} showed that by varying the polarization and the delay between two ultrashort laser pulses, one can induce unidirectional molecular rotation, thereby forcing molecules to rotate clockwise or counterclockwise under field-free conditions. This phenomenon was given both classical and quantum interpretation in  Ref.~\cite{KhodorkovskyPRA11b}.
 It was suggested that collisions in an ensemble of such unidirectionally rotating molecules lead to macroscopic vortices \cite{SteinitzPRL12}.
 A chiral pulse train -- a sequence of linearly polarized pulses with the polarization direction rotating from pulse to pulse by a controllable angle -- was also used to achieve selectivity and directionality of laser-induced rotational excitation~\cite{ZhdanovichPRL11, FlossPRA12, BloomquistPRA12}. In a striking development, Fleischer~\textit{et al.}~\cite{FleischerPRA06, FleischerPRL07} showed the possibility to achieve a nuclear-spin-selective molecular alignment for the case of N$_2$ molecules composed of $^{15}$N and $^{14}$N atoms. Following this work, Gerschnabel and Averbukh~\cite{GershnabelPRA08} presented a theoretical study of the spin-selective alignment for ortho- and para- water isomers. Orienting molecules is important for studying fs-time-resolved angular and momentum distributions  of photoelectrons  from large molecules~\cite{HolmegaardNatPhys10, HansenPRL11, HansenJPB11}.  A three-dimensional photoelectron momentum distribution from naphthalene molecules fixed in space was observed by Maurer~\textit{et al.}~\cite{MaurerPRL12}.  Pabst~\textit{et al.}~\cite{PabstPRA10} presented a theory illustrating the feasibility to extract information on  molecular structure from x-ray scattering from 3D aligned molecules, also see Ref.~\cite{FilsingerPCCP11}. Time-resolved studies of electronic-vibrational dynamics during a photochemical reaction of CS$_2$ molecules fixed in space was reported by Bisgaard~\textit{et al.}~\cite{BisgaardSci09}.
 Averbukh and Arvieu~\cite{AverbukhPRL01} studied semiclassical catastrophes in the dynamics of a molecular rotor driven by a strong time-dependent field, and provided a recipe for the rotational squeezing by a sequence of laser pulses. Short-pulse rotational coherence spectroscopy can also be used to probe molecular structure, yielding information on molecular properties such as rotational constants and polarizabilities of large molecules and clusters, for both the ground and the electronically excited states~\cite{OhlineLC94, RiehnCP02, MatylitskyJCP03}. Molecules interacting with short laser pulses can serve as a model to investigate nonlinear quantum phenomena in the dynamics of a kicked rotor, such as Anderson-like localization in angular momentum space~\cite{FlossPRA12b}. Quantum resonances in the spectrum of a kicked rotor were investigated experimentally and theoretically in Ref.~\cite{ZhdanovichPRL12}.  It was also shown that diatomic molecules in the presence of magnetic~\cite{SchmelcherPRA90} and laser~\cite{MoiseyevJPB08, HalaszJPCA12, HalaszCP12} fields possess conical intersections, which in the field-free regime are present only in molecules consisting of more than two atoms. This opens up new possibilities to study nonadiabatic effects in the vicinity of conical intersections. Finally, it was proposed to use short laser pulses to map out the vibrational wavefunction of diatomic molecules~\cite{LemFriPRL09, LemFriPRArapid09} and fine-tune the molecular vibrational levels~\cite{LemFriJPCA10}. 
 It was shown that far-off-resonant light can be used to tune shape resonances in atom-atom scattering and thereby enhance the photoassociation yield~\cite{RuzinArxiv11, GonzalezFerezPRA12}.




\section{Manipulating molecules by non-conservative forces}
\label{sec:cooling}

Resonant scattering of laser light by atoms or molecules is an inherently dissipative process due to the stochastic nature of spontaneous emission. This process is at the core of laser cooling, a versatile cooling technique for atoms possessing a cycling transition~\cite{MetcalfBook}. Unlike atoms, molecules lack closed optical transitions required for efficient laser cooling. When returning to the electronic ground state by emitting radiation, electronically excited molecules are destined to populate a range of vibrational levels. This happens primarily because electronic excitation generally changes the strength of the chemical bond, resulting in a displacement of the minimum of the corresponding potential energy curve. Despite the lack of cycling transitions, several avenues for laser cooling of molecules have been explored, starting from the work of Bahns, Stwalley, and Gould~\cite{BahnsJCP96}.

In 2004, di~Rosa \cite{DiRosaEPJD04} argued that certain diatomic molecules have the electronic transitions that are quasi-cycling. However, electronic transitions, even if cycling between the same vibrational states, may still lead to spreading of the population over rotational states due to the dipole selection rules. Stuhl~\textit{et al.}~\cite{StuhlPRL08} demonstrated the possibility of closing the rotational transitions $J\to J'$ by choosing the ground-state rotational angular momentum, $J$, to be larger than the angular momentum in the excited electronic state one, $J'$. They showed the feasibility of laser cooling of TiO and identified the class of molecules that are exceptionally good candidates for laser cooling -- those without hyperfine structure and having good Franck-Condon overlaps.  An example of such molecules is  SrF($^2\Sigma$). The unpaired electron of this molecule does not participate much in chemical bonding. Therefore, the electronic excitation of SrF does not change dramatically the potential energy so the vibrational ground level of the excited electronic state preferentially decays to the vibrational ground level of the electronic ground state.  As in any molecule, there is small leakage of population to other vibrational levels. However, as demonstrated by  Shuman and coworkers~\cite{ShumanNature10}, this can be fixed by adding a few repumping lasers, collecting all the leaked population back to the vibrational ground level of the electronic excited state. This ground-breaking experiment showed that laser cooling of molecules is indeed possible, opening also the prospects for manipulation of molecular beams with dissipative laser forces ~\cite{ShumanPRL09, BarryPRL12}. 

Building on the theoretical and experimental work mentioned above, several authors considered the possibility of laser cooling of various molecules such as TlF~\cite{HunterPRA12},  RaF~\cite{IsaevPRA10}, and YbF~\cite{TarbuttNJP13}. Hummon~\textit{et al.} demonstrated laser cooling and magneto-optical trapping of YO molecules in two dimensions~\cite{HummonPRL13}. However, the range of molecules amenable to the laser cooling technique demonstrated by Stuhl, Shuman, Hummon, and their coworkers remains rather narrow. 
The development of a general laser cooling method applicable to any diatomic as well as polyatomic molecule is a challenging problem currently researched by many groups. Most notably, it was suggested that the detrimental decay processes can be damped by coupling  a molecule to an optical or microwave cavity, which might allow for cavity cooling of external and internal molecular motion~\cite{VuleticPRL00, MorigiPRL07, KowalewskiAPB07, LevPRA08, WallquistNJP08, KowalewskiPRA11}. Averbukh and Prior~\cite{AverbukhPRL05, VilenskyPRA06} proposed an alternative method for laser cooling of molecules by an optical shaker -- a standing wave of far-off-resonant  light that exhibits sudden phase jumps, whose value is controlled by the feedback loop. In such a way the cooling scheme combines the principles of Sisyphus~\cite{DalibardJOSAB89} and stochastic~\cite{MohlPhysRep80} cooling. Vilensky~\textit{et al.} proposed to use a bistable optical cavity for a Sisyphus-type cooling, where the cavity undergoes sudden transitions between two energy states~\cite{VilenskyPRL07}. These proposals are currently awaiting experimental realization.

Optical pumping is routinely used to prepare atoms in a particular quantum state. Extending this technique to molecules is impeded by the same problem of multi-level structure that has haunted the development of laser cooling of molecules. Only recently, it was shown that diatomic molecules can be cooled, vibrationally and rotationally, to the ground rovibrational level using optical pumping~\cite{ViteauSci08, SofikitisNJP09, ManaiPRL12}. These experiments exploit broadband lasers, generating femtosecond pulses  shaped to remove the frequency band that would excite the ground state.  Without this frequency, the laser pulses redistribute the population leading to efficient accumulation of population in the ground state.

In a remarkable recent work, an opto-electrical method of cooling molecules was proposed and realized by Zeppenfeld~\textit{et al.}~\cite{ZeppenfeldPRA09, ZeppenfeldNature12}. 
Closely related to the idea of  single-photon molecular cooling~\cite{NareviciusNJP09},
this method combines the techniques of Stark deceleration and laser cooling by applying an electric-field potential that generates substantially different Stark shifts in two molecular states. The energy is removed by letting the molecules climb the field gradient in the state with a bigger Stark shift and return in a state with a smaller Stark shift. The entropy is removed by a cycling transition between the states, involving spontaneous emission. Due to the difference in the Stark shifts, the cycling transition leads to a large Sisyphus effect, with much kinetic energy taken away in each cycle. This allows for cooling using only a few instances of the spontaneous emission, bypassing the population leakage problem mentioned above.

Dissipation due to the near-resonant scattering of light can be turned into a useful resource for quantum state preparation~\cite{DiehlNatPhys08, VerstraeteNatPhys09, Weimer2010, BarreiroNat11}. Lemeshko and Weimer showed that using engineered dissipation it is possible to generate metastable bonds between atoms or molecules, thereby extending the notion of `bonding' from purely conservative to dissipative forces. The bond arises due to the interaction-dependent coherent population trapping and manifests itself as a stationary state of the scattering dynamics that confines the atoms or molecules at a fixed distance from each other. Remarkably, the dissipative bonding appears possible even for atoms and molecules interacting purely repulsively~\cite{LemWeimDiss}.

\vspace{1cm}

\section{Controlled molecular collisions}
\label{sec:MolInteractions}

\subsection{Towards ultracold molecules}

Using electromagnetic fields to control intermolecular interactions is a long-standing goal in molecular physics.  Of particular importance to chemical physics has been the effort directed towards external field control of molecular collisions, whether elastic, inelastic, or chemically reactive. 
At ambient temperatures, molecules reside in a manifold of internal states and move with a wide range of velocities and angular momenta. It is this wide distribution of accessible internal and motional states, each generally leading to different reaction events, that makes external field control of bi-molecular interactions in a thermal molecular gas extremely difficult, if not impossible. These complications are removed when molecules are cooled to subKelvin temperatures.  
The experiments on cooling molecules and producing ultracold molecules from ultracold atoms have thus opened another research avenue in molecular physics, exploiting {\it controlled few-molecule dynamics}.

When cooled to sufficiently low temperatures, molecules can be confined in magnetic, electric or optical traps \cite{CarrNJP09}, as described in Section~\ref{sec:trapping}. This effectively prepares molecular ensembles in a single (or a small number of) field-dressed states in a non-perturbative regime, where the molecule-field interactions are more significant than the energy of the translational motion. When the first experiment on magnetic trapping of a molecular radical CaH was reported \cite{WeinsteinCaH}, little was known about the interaction properties of molecules in this regime of molecule-field interactions. The experiments on cooling molecules -- and the promise of marvelous applications described in Ref. \cite{CarrNJP09} -- spurred intense research on binary molecular interactions in strong dc fields. Most of this research was motivated by three general questions: Are molecules in a particular field-dressed state stable against collisional interactions? If not all molecules are, which are and which are not? To what extent do external fields affect intermolecular interactions at low temperatures? These questions are particularly relevant for experiments with molecules in dc magnetic or electric traps, because dc fields, by the Earnshaw's theorem \cite{Earnshaw1839, WingPRL80}, always confine particles in excited Zeeman or Stark levels, leaving an opportunity for molecules to decay to lower energy levels and exit the trap.

The figure of merit used to quantify the answer to the first question is the ratio of cross sections for elastic scattering (responsible for translational energy thermalization within a trap) to cross sections summed over all possible inelastic and reactive scattering channels (leading to trap loss). This elastic-to-inelastic ratio, usually denoted by $\gamma$, indicates if a molecular ensemble can be cooled in an external field trap ($\gamma > 1000$) or not ($\gamma < 100$). 
In a zealous effort to find mechanisms for enhancing $\gamma$ to large values, multiple theoretical studies explored the possibility for modifying elastic, inelastic, and reactive collisions of molecules by external fields.  The pioneering study of Volpi and Bohn \cite{VolpiBohnPRA02} showed that Zeeman transitions (or Stark transitions, for that matter) in collisions of ultracold molecules must be suppressed at low external fields.  The suppression occurs because any Zeeman or Stark transition in ultracold $s$-wave collisions must be accompanied by a change of orbital angular momentum for the relative motion of the colliding molecules, which leads to centrifugal barriers in outgoing collision channels. This phenomenon can also be interpreted in terms of the energy dependence of the cross sections for Zeeman or Stark transitions in the limit of zero external field \cite{KremsPRA03}.  These results imply that any (chemically non-reactive) molecules, if cooled to a low enough temperature to be trappable in a vanishingly shallow trap, must be stable. This is encouraging because it indicates that any molecules that do not chemically react can be cooled by evaporation in a dc magnetic or electric trap, if the starting temperature of the molecular ensemble is cold enough. However, the problem is that several calculations \cite{AvdeenkovPRA02, AvdeenkovPRA05,LaraPRL06,TscherbulNJP09} showed that large values of $\gamma$ can generally be reached only when the trapping fields are so small that the starting temperature must be very low ($< 1$ $\mu$K is a good ballpark limit).

As described in Section~\ref{sec:beams} , there are several different experimental techniques that can be used to prepare molecules at temperatures about $1$ mK. However, bridging the gap between $1$ mK and 1$\mu$K has proven difficult. A number of authors \cite{VolpiBohnPRA02, KremsPRA03, KremsPRA03b, BalakrishnanJCP03, KremsDalgarnoJCP04, CybulskiJCP05, TscherbulPRL06, TsherbulKremsJCP2006, LaraPRL06, TscherbulPRA07, AbrahamssonJCP07, GonzalezMartinezPRA07, ZuchowskiPRA09,  SoldanFD09, CampbellDoylePRL09, HutsonPRL09, WallisPRL09, ZuchowskiPCCP11, JanssenPRA11, WallisEPJD11, SkomorowskiJCP11, TokunagaEPJD11, GonzalezMartinezPRA11, WallisPRA11, SuleimanovJCP12, BalakrishnanCPL97, BalakrishnanCPL97b, BalakrishnanPRL98, ForreyPRA98, ForreyPRA99, BalakrishnanJCP00, BalakrishnanJPCA01} researched the dynamics of molecular collisions in this temperature interval in order to understand the prospects for cooling molecules from mK temperatures to ultracold temperatures below 1 $\mu$K. 
Many of the early studies were focussed on the dynamics of molecules in a helium gas \cite{VolpiBohnPRA02, KremsPRA03, KremsPRA03b, BalakrishnanJCP03, CybulskiJCP05, TscherbulPRL06, TsherbulKremsJCP2006,  TscherbulPRA07, GonzalezMartinezPRA07}. Partially motivated by the successful experiments on magnetic trapping of molecules in a helium buffer gas \cite{WeinsteinCaH, deCarvalhoEPJD99, EgorovPRA02, EgorovEPJD04, CampbellPRL07, HummonPRA08, CampbellDoylePRL09, TsikataNJP10, HummonPRL11}, these papers also argued that if collisions with weakly interacting helium atoms are sufficiently strong to destroy trapped samples, evaporative cooling of molecules in the absence of a buffer gas must be impossible. The results predicted that diatomic molecules in $\Sigma$ electronic states with large rotational constants should be stable in a magnetic traps in the environment of He buffer gas and that molecules in non-$\Sigma$ electronic states must prefer to undergo inelastic scattering generally leading to low values of $\gamma$. While the former prediction was confirmed in a number of experiments \cite{CampbellPRL07, HummonPRA08, CampbellDoylePRL09, TsikataNJP10, HummonPRL11}, the latter was -- fortunately! -- proven incorrect in a recent startling experiment by Stuhl {\it et al.}~\cite{StuhlNature12}. As the experiments of  Ref.~\cite{StuhlNature12} and the theory by 
Bohn and coworkers \cite{AvdeenkovBohnPRL03, AvdeenkovPRA04, TicknorBohnPRA05, AvdeenkovPRA06, BohnArxiv13} show, there is a general mechanism that may suppress inelastic collisions of molecules with a $\Lambda$-doubled structure placed in superimposed electric and magnetic fields. Combined with 
large cross sections for elastic scattering \cite{TicknorBohnPRA05}, this may make the creation of a Bose-Einstein Condensate of $\Pi$-state molecules possible! 

Another possibility for bridging the $1$ mK -- 1 $\mu$K temperature gap is to immerse a gas of molecules into a buffer gas of already ultracold atoms. The calculations by Hutson and coworkers \cite{SoldanFD09, GonzalezMartinezPRA11, WallisPRA11} demonstrated that alkaline-earth metal atoms, being structureless and therefore weakly interacting, can be used for sympathetic cooling of molecules such as NH. Interestingly, Tscherbul and coworkers \cite{TscherbulPRA11} have recently found that sympathetic cooling may work even with alkali metal atoms, such as Li, despite their propensity to generate strongly attractive and anisotropic interactions with molecules. 
This implies that certain molecules, particularly $\Sigma$-state open-shell molecules with weak fine-structure interactions, may be generally amenable to sympathetic cooling at $T \sim 1$ mK, which was confirmed by preliminary calculations for NH--NH scattering in magnetic field \cite{JanssenPRA11}.  Following the theoretical work by Zuchowski and Hutson \cite{ZuchowskiPRA09}, 
Parazzoli~\textit{et al.} \cite{ParazzoliPRL11} reported an experimental study of the effects of electric fields on low temperature collisions of ND$_3$ molecules with Rb atoms. Unfortunately, the results reveal high probability of inelastic scattering of molecules in the low-field seeking states, indicating that sympathetic cooling of symmetric top molecules in a buffer gas of alkali metal atoms is likely unfeasible. Following the accurate calculation of the global potential energy surface for the NH--NH collision system in the 
quintet spin state \cite{JanssenJCP09}, several authors performed quantum scattering calculations for NH--NH collisions in a magnetic field \cite{JanssenPRA11, SuleimanovJCP12}. The results ignoring non-adiabatic couplings to lower spin states of the two-molecule complex suggested that inelastic scattering in collisions of NH molecules must be generally insignificant due to the small magnitude of the spin-spin interaction in the molecule. However, a more refined calculation by Janssen, van der Avoird and Groenenboom \cite{JanssenPRL13} indicated that the non-adiabatic couplings may preclude the evaporative cooling. 
Other, highly promising methods for cooling molecules to ultracold temperatures have been recently proposed and demonstrated. For example, the groups of Zoller and B{\"u}chler showed that the translational energy of molecules can be removed by coupling the molecular ensemble to a gas of Rydberg atoms and tuning the atom-molecule interactions with external fields~\cite{ZhaoPRL12, HuberPRL12}. Zeppenfeld and coworkers showed that electrically trapped molecules can be efficiently cooled by a combination of optical 
and electric field forces \cite{ZeppenfeldPRA09, ZeppenfeldNature12}. It is thus foreseeable that a  great variety of diatomic, and perhaps polyatomic, molecules can be cooled to ultracold temperatures directly from a thermal gas in a near future~\cite{LiJCP136}. At present, the experiments with ultracold molecules are limited to diatomic molecules produced from ultracold atoms. 

While looking for mechanisms to enhance $\gamma$, the theoretical studies have found several mechanisms for controlling molecular interactions at both cold ($<$ 1 K) and ultracold ($< 1$ mK) temperatures. For example, Tscherbul and coworkers \cite{TscherbulPRL06, TsherbulKremsJCP2006, AbrahamssonJCP07}  found that the dynamics of magnetic Zeeman relaxation in collisions of $^2\Sigma$ and $^3\Sigma$ molecules is very sensitive to external electric fields, which affect the rotational structure and modify the fine-structure interactions inducing spin-changing transitions. Combined electric and magnetic fields can also be used to create molecules near avoided crossings between Zeeman levels arising from rotational states of different symmetry, making them extremely sensitive to small variations of external (magnetic or electric) fields \cite{FriHerPCCP00, BocaFri00}.  Molecules in a $^2\Pi$ electronic state exhibit an interesting dependence of the collision cross sections on the electric field strength, resulting from the shifts of the asymptotic collision channels and other, less understood, phenomena \cite{TscherbulKremsFD09}. As pointed out by Bohn and coworkers  \cite{AvdeenkovBohnPRL03, AvdeenkovPRA04, TicknorBohnPRA05, AvdeenkovPRA06, BohnArxiv13}, dipole-dipole interactions lead to the appearance of avoided crossings in the interaction potential of the molecule-molecule collision complex. These avoided crossings can be tuned by an external electric field, leading to dramatic changes in the collision dynamics at ultracold temperatures. By changing the internal structure of molecules, external fields induce molecule-molecule scattering resonances \cite{AvdeenkovBohnPRL03, AvdeenkovPRA04, TicknorBohnPRA05, AvdeenkovPRA06, TscherbulNJP09}. Stuhl~\textit{et al.} experimentally studied the avoided crossing in an OH molecule in combined electric and magnetic fields~\cite{StuhlPRA12} which ultimately allowed to evaporatively cool OH~\cite{StuhlNature12}. 
In a remarkable paper \cite{TicknorPRL08}, Ticknor showed that two-body scattering properties of dipolar species in the limit of strong dipole-dipole interactions and at ultracold temperatures are universal functions of the dipole moment, mass of the colliding species and collision energy. 
Bohn, Cavagnero and Ticknor \cite{BohnNJP09} showed that the universality extends to a wider range of energies and studied the deviations from universality, which can provide information about the short-range part of the intermolecular interaction potentials.


\subsection{Towards controlled chemistry}

Chemical dynamics of molecules in external fields was first studied in a rigorous quantum calculation by Tscherbul and Krems \cite{TsherbulKremsJCP2006}. Beyond demonstrating the feasibility of such calculations, this work showed that the probabilities for chemically reactive events, just like the cross sections for elastic and inelastic scattering, may be quite sensitive to external electric fields. 
Controlling  chemical interactions of molecules at subKelvin temperatures can be used as a tool to study the role of fine and hyperfine interactions as well as non-adiabatic effects in elementary chemical processes.  For example, confining molecules in a magnetic trap leads to co-alignment of their magnetic moments, which restricts the adiabatic interaction between the molecules to the maximum spin state. Since large spin states are usually characterized by significantly repulsive exchange interactions, chemical reactions of spin-aligned molecules are generally very slow, especially at low temperatures \cite{KremsPCCP08}. The non-adiabatic interactions may serve to enhance the chemical reactivity by inducing transitions to lower spin states of the reaction complex \cite{JanssenPRL13}. Tuning the non-adiabatic interactions with external fields, for example as suggested in Ref. \cite{TscherbulPRL06}, can be used as a tool to quantify the non-adiabatic interactions. 

Molecular beam technique can serve as a basis to study cold reactive and nonreactive collisions~\cite{HerschbachFD09}.  Van de Meerakker and coworkers studied rotationally inelastic scattering of state-selected and velocity controlled OH molecules with D$_2$ molecules and a wide range of rare-gas atoms~\cite{GilijamseSci06, ScharfenbergPCCP10, KirstePRA10, ScharfenbergPCCP11, ScharfenbergEPJD11, GubbelsJCP12},  as well as with state-selected  NO molecules~\cite{KirsteSci12}. Eyles~\textit{et al.} performed detailed studies of differential cross sections in rotationally inelastic NO--Ar collisions~\cite{EylesNatChem11}. Strebel~\textit{et al.} studied elastic collisions of a beam of SF$_6$, decelerated using a rotating nozzle, with trapped ultracold Li atoms~\cite{StrebelPRA12}.  Ohoyama and coworkers studied collisions between polyatomic molecules oriented by an electric hexapole with molecules and atoms aligned  by a magnetic hexapole~\cite{WatanabePRL07, MatsuuraJPCA11, OhoyamaJCP12}.

The efforts to study cold collisions have long been frustrated by the need to create two slow beams of high enough intensity. An alternative technique, based on merged molecular beams, allows one to study cold collisions between molecules moving fast in the laboratory frame, provided the velocity distribution in the beams is narrow enough. Henson~\textit{et al.} used the merged-beam technique to study the Penning-ionization reaction He$^\ast +$H$_2 \to$ He$+$H$_2^+$ at sub-Kelvin temperatures~\cite{HensonSci12}. Scheffield~\textit{et al.} developed a pulsed rotating supersonic source for use with merged molecular beams \cite{SheffieldRSI12}. Wei~\textit{et al.} discussed the prospects of the merged beam technique in Ref.~\cite{WeiJCP12}.

Another way to study cold collisions is to confine one or both of the collision partners in a trap. Sawyer~\textit{et al.} studied collisions of trapped OH molecules with supersonic beams of He and D$_2$~\cite{SawyerPRL08}, and with a velocity selected beam of ND$_3$~\cite{SawyerPCCP11}. Parazzoli~\textit{et al.} \cite{ParazzoliPRL11} merged separately trapped ND$_3$ molecules and Rb atoms, in order to study the electric field effect on the cross sections. In a contribution to this special issue Stuhl~\textit{et al.} studied the effect of an electric field on rotationally inelastic scattering of OH molecules~\cite{StuhlMolPhys13}.

Ultracold chemistry has become real with the creation of a dense ensemble of ultracold KRb molecules \cite{OspelkausNatPhys08, NiScience08}. By measuring the trap loss of KRb molecules, the experiments demonstrated many unique features of chemical reactions in the limit of zero temperature, such as the effect of quantum statistics and external electric fields on chemical encounters~\cite{NiJinYeNature2010, OspelkausScience10, MirandaYeJin11}. 
While exciting, ultracold reactions are also detrimental to the progress of experiments aimed at the creation of a quantum degenerate gas of ultracold molecules. The experimental studies were therefore accompanied by many calculations seeking to suppress chemical reactivity of ultracold molecules by external fields \cite{MicheliPRL10, IdziaszekPRL10, QuemenerPRA10, IdziaszekPRA10, MeyerPRA10, QuemenerPRA11, QuemenerPRA11b, JuliennePCCP11, MaylePRA13}. One possibility, it was found, is to confine molecules in low dimensions by optical lattices and apply an electric field. 

Following the pioneering work of Petrov and coworkers \cite{PetrovPRA03}, the effects of dimensionality on inelastic collisions and chemical reactions was studied by Li and Krems \cite{LiPRA09}. The results showed that confining molecules in a quasi-2D geometry leads to enhanced values of $\gamma$. The work reported in Refs. \cite{NiJinYeNature2010, OspelkausScience10, MirandaYeJin11} showed that for polar molecules in a quasi-2D gas this enhancement can be dramatically increased by applying an electric field perpendicular to the plane of confinement. The electric field serves to align the molecules, thereby creating long-range repulsive barriers stimulated by the dipole-dipole interactions. 
Chemical reactions in such a gas are determined by the rate of tunneling through the long-range barriers~\cite{LipoffMolPhys10}, which as shown earlier by Bohn and coworkers, is sensitive to external fields.   
Even more effectively, the long-range repulsive barriers can also be created by dressing molecules with a combination of dc electric and microwave fields \cite{MicheliPRA07, GorshkovPRL08b}, and far-off resonant laser fields~\cite{LemeshkoPRA11Optical, LemFri11OpticalLong}. 
As was shown in refs. \cite{BuchlerZollerPRL07, MicheliPRA07, GorshkovPRL08b, GuidoBretBook}, this technique can be used to create a gas of molecules with repulsive long-range interactions that completely preclude the molecules from penetrating to the region of short-range interactions, which, under appropriate conditions, can lead to the formation of self-assembled dipolar crystals and other interesting many-body effects discussed in the following section. 

Trapped ions represent a new paradigm in the study of cold chemistry ~\cite{RothPRA06, WillitschPRL08, BellFarDiss09, HallPRL12, TongCPL12}. Atomic or molecular ions bombarded by neutral species, coming, for example, from controlled molecular beams, allow for an extremely long interrogation time. This opens the possibility of measuring the reactions of a single ion ~\cite{StaanumPRL08, HansenAngChem12, RatschbacherNature12} or reactions with very low rates. Since ions are trapped, external field control of molecular beams can be used to study ion-molecule chemistry with high energy resolution, as was attempted in experiments on inelastic collisions of molecules in Stark decelerated beams with trapped atoms \cite{SchlunkPRL07, MarianEPJD10}. The advantage of using ions is that the charged products of a chemical reaction can also be trapped, which allows one to measure the reaction product state distributions~\cite{TongCPL12}.


Chemical dynamics of molecules at ambient temperatures can also, in principle, be controlled by aligning molecules or by preparing molecules in superpositions of different electronic states.  Although studying molecular collisions in fields at elevated temperatures is of a substantial interest due to chemical applications~\cite{LevineBook}, quantum mechanical treatment of such collisions is challenging due to a large number of states involved in the dynamics. A quasi-classical trajectory method has been used to elucidate the effect of pendular orientation of DCl  on its reaction with H atoms~\cite{AoizCPL98} and the effect of an intense infrared laser field on the H$+$H$_2$ exchange reaction rate~\cite{IvanovCPL96}. On the other hand, simple analytic models of molecular collisions, essential for understanding the underlying physics, are scarce and mostly limited to the Wigner regime of very low kinetic energy~\cite{SadeghpourJPB00}. Lemeshko and Friedrich developed an analytic model of molecular collisions in fields and applied it to the study of the scattering dynamics~\cite{LemFriJCP08, LemFriIJMS09, LemFriJPCA09, LemFriPRA09} and stereodynamics~\cite{LemFriPCCP10, LemJamMirFriJCP10} at thermal collision energies, as well as to multiple scattering of matter waves~\cite{LemFriPRA10}. The model's accuracy increases with collision energy; it has no fitting parameters and remains analytic in the presence of external fields.

\section{Controlling many-body phenomena}
\label{sec:many-body}

The exquisite control over the internal and motional dynamics of molecules at low temperatures paves the way for creating unique many-body systems with interesting properties \cite{GorshkovPRL08, BuchlerZollerPRL07, FregosoPRB10, BaranovPhysRep08, LahayePfauRPP2009}. For example, several studies, exploiting the analogy with optical shielding of interactions in ultracold atomic gases \cite{WeinerRMP99}, showed that the dipole-dipole interactions between polar molecules confined in a quasi-2D geometry by an optical lattice and subjected to microwave fields can be made repulsive \cite{BuchlerZollerPRL07, MicheliPRA07, GorshkovPRL08, ArtamonovPRL12}.  This leads to the formation of self-assembled dipolar crystals. Molecular dipolar crystals have been proposed as high-fidelity quantum memory for hybrid quantum computing involving an interface of a solid-state device and trapped molecules \cite{RablPRL06}.  
B{\"u}chler and coworkers extended this work to show that a suitable combination of microwave and dc electric fields can be used to engineer three-body interactions in a molecular gas \cite{BuchlerNatPhys07}, which, when trapped on an optical lattice, realizes the Hubbard model with strong three-body interactions. The possibility to form fermonic molecules in an optical lattice and spectroscopy of such a system has been theoretically investigated in refs.~\cite{HuPRA11, HazzardPRA11}.

Trapping ultracold  non-polar and polar molecules in an optical lattice, recently demonstrated in a number of  experiments \cite{ChotiaPRL12, DanzlNature10, StellmerPRL12, ReinaudiPRL12}, is a major step towards quantum simulators of condensed matter models. Several recent papers showed 
that ultracold molecules confined on an optical lattice can be used to realize a wide range of model many-body Hamiltonians \cite{BaranovPhysRep08, LahayePfauRPP2009, BarnettPRL06, DallaTorrePRL06, WallNJP09, PolletPRL10, SunPRB10, DalmontePRL10, Gorshkov11b, TrefzgerJPB11, KunsPRA11, MikelsonsPRA11, DalmontePRL11, BabadiPRA11, KnapPRB12, ManmanaPRB13}, including extended Hubbard models with long range interactions \cite{OrtnerNJP09, WallMPRA10, BarbaraPRL10, BhongalePRL12, WallArxiv12}, a variety of lattice spin models \cite{MicheliNatPhys06, BrennenNJP07, SchachenmayerNJP10, PeresRiosNJP10,  HerreraPRA10, gorshkovPRA11, ZhouPRA11, LemeshkoPRL12, GorshkovMolPhys13, WeimerMolPhys13, WallArxiv13, HazzardPRL13}, and polaron models \cite{HerreraPRA11, HerreraPRL13}. Yao~\textit{et al.} proposed a scheme to realize topological flatbands~\cite{YaoPRL12} and fractional Chern insulators~\cite{YaoPRL13} in dipolar gases; Kestner~\textit{et al.} predicted a topological Haldane liquid phase in a lattice with cold molecules \cite{KestnerPRB11}. These studies exploit the rich structure of molecules enabled by the rotational, spin and hyperfine states and the possibility of  tuning the couplings between these degrees of freedom as well as the dipole-dipole interactions between molecules by externally applied fields.  In many instances, it is possible to apply a combination of dc and microwave fields such that the individual molecules exhibit isolated field-dressed states with particular features identical to those of the spin degrees of freedom in lattice-spin Hamiltonians. The external fields can then be tuned to explore the phase diagram of the many-body system on the lattice. 
  Of particular interest is the possibility of creating quantum phases with topological order
\cite{ManmanaPRB13}, resilient vis a vis perturbations preserving the topology. Recently Yan~\textit{et al.} reported the first experimental realization of a spin model with KRb molecules on a 3D optical lattice~\cite{YanArxiv13}.

A significant focus of research with ultracold atomic gases in the past decade has been on emergent phenomena, such as solitons, rotons, vortices, spin waves, and polarons. Molecules offer new degrees of freedom that can be used to explore new regimes of collective phenomena \cite{RonenPRA06, RonenPRA06b, BortolottiPRL06, RonenPRL07, RonenPRA07, WilsonPRL08,  BohnLaserPhysics2009, WilsonPRA09, WilsonPRA09b, WallMPRA10, WilsonPRL10, RonenPRA10, TicknorPRL10, WilsonPRA11, ZillichPCCP11, WilsonNJP12, WilsonPRA12, WallArxiv12}. The possibility of mixing different parity states with moderate electric fields enables new interactions that give rise to unique features of collective dynamics in many-body molecular systems. For example, collective rotational excitations of molecules trapped in an optical lattice can pair up forming Frenkel biexcitons \cite{XiangPRA12}, something that cannot occur in natural solid-state systems with inversion symmetry. The ability to control intermolecular interactions can also be used to engineer many-body, nanoscopic systems allowing for controlled energy transport \cite{XiangArxiv12}, as well as open quantum systems with tunable coupling to the bath \cite{HerreraPRA11}.  While most of the work mentioned above is focused on polar molecules with tunable dipole-dipole interactions, the many-body behaviour of non-polar molecules can also be controlled by tuning the quadrupole-quadrupole interactions. The peculiar symmetry and broad tunability of the quadrupole-quadrupole couplings results
in a rich phase diagram featuring unconventional BCS and charge density wave phases, and opens up the prospect to create a topological superfluid~\cite{BhongalePRL13}. Quadrupolar particles, such as homonuclear molecules or metastable alkaline-earth atoms are  currently available in experiments at higher densities compared to polar molecules. The coldest of the currently available molecules, Cs$_2$, can be prepared in optical lattices close to unit filling at temperatures significantly below the photon recoil energy of 30~nK~\cite{DanzlNature10, NagerlPrivate}.

\vspace{1cm}

\section{Entanglement of Molecules and Dipole Arrays}

Applications of the concepts of quantum information theory are usually related to the
quantum mechanical effects of superposition, interference, 
and entanglement \cite{Entag-Amico,Entag-Osborne,Entag-Kais-2007,mazz07}.  For decades, theoretical
chemists have encountered and analyzed these quantum effects from the point of view
of bonding. Combining results and insights of quantum information science
with those of chemical physics may shed new light on the dynamics of the chemical bond.
Researchers from the quantum information and chemistry communities are already
converging upon several questions. As an example, recent ultrafast experiments on excitonic migration
in photosynthetic complexes and polymers have shown long-lived
coherences on the order of hundreds of femtoseconds~\cite{LambertNatPhys12}.

Since the original proposal by DeMille~\cite{Entag-DeMille}, arrays of ultracold polar molecules have
been counted among the most promising platforms for the implementation of a quantum
computer \cite{Entag-Yelin, krems09}. The qubit of such an array is realized by a single dipolar
molecule entangled  with the rest of the array's molecules via the dipole-dipole interaction.
Polar-molecule arrays appear as scalable to a large number of qubits as neutral-atom
arrays do, however the dipole-dipole interaction furnished by polar molecules offers a faster
entanglement, one resembling that mediated by the Coulomb interaction for ions. At
the same time, cold and trapped polar molecules exhibit similar coherence times as
those encountered for trapped atoms or ions. The first complete scheme proposed for
quantum computing with polar molecules was based on an ensemble of ultracold polar
molecules trapped in a one-dimensional optical lattice, combined with an inhomogeneous
electrostatic field. Such qubits are individually addressable, due to the Stark shift
which is different for each qubit in the inhomogeneous electric field~\cite{Entag-Viv}.

A subsequent proposal has shown that it should be possible to couple polar molecules
into a quantum circuit using superconducting wires \cite{Entag-68}. The capacitive, electrodynamic
coupling to transmission line resonators was proposed in analogy with coupling to
 Rydberg atoms and Cooper pair boxes~\cite{Entag-69}. Compatibility with the microwave circuits is
ensured by the frequencies of the transitions between molecular Stark states, 
which occur typically in the rf range. The coupling of polar molecules to microwave striplines
carries along the following advantages: first, it allows for detection of single molecules
by remote sensing of transmission line potentials as well as for efficient quantum-state
readout; second, the molecules can be further cooled by microwave spontaneous emission
into on-chip transmission lines; and finally, the coupling to the strip-line entangles the
molecules and thus enables nonlocal operations. Addressability is achieved by local gating with electrostatic
  fields and single-bit manipulations can be accomplished by using
local modulated electric fields.

A pure state of a pair of quantum systems is called entangled if it is
unfactorizable, as for example, the singlet state of two spin-$1/2$ particles,
a mixed state is entangled if it cannot be represented as a mixture of factorizable pure
states~\cite{Entag-Kais-2007}.  Study of entanglement can be done by calculating
the pairwise concurrence~\cite{Entag-Wootters}, which is a good measure of entanglement, for one, two,
and three-dimensional arrays of trapped dipoles, mutually coupled by the dipole-dipole
interaction and subject to an external electric field. 
Quantum entanglement can also be studied using trapped polar molecules.
Arrays of polar molecules can be prepared in optical lattices with full control over the
internal states including the hyperfine structure. Recently, Qi \textit{et al.} 
considered using rotational states of polar linear \cite{Entag-Qi1,Entag-Qi3} and symmetric-top~\cite{Entag-Qi2} molecules as qubits 
  and  evaluated entanglement of the 
 pendular qubit states for two linear dipoles, characterized by pairwise concurrence,
  as a function of the molecular dipole moment and rotational constant, strengths of the external 
  field and the dipole-dipole coupling, and ambient temperature. 
  In principle, such weak entanglement can be sufficient for operation of logic gates, provided the resolution is high enough to detect the state shifts unambiguously~\cite{Entag-Qi1}. In practice, however, for many candidate polar molecules it appears a challenging task to attain adequate resolution. In a subsequent study, the authors considered 
  symmetric top molecules.  The latter offer advantages resulting from a first-order Stark effect, 
 which renders the effective dipole moments nearly independent of the field strength. 
 For a particular choice of qubits, the electric dipole interactions 
  become isomorphous with NMR systems for which many techniques enhancing logic
gate operations have been developed~\cite{Entag-Jing}.

Entanglement can be found in strongly
interacting atomic and molecular gases, however it is challenging to generate highly
entangled states between weakly interacting particles in a scalable way. Based on the work of Lemeshko and Friedrich~\cite{LemeshkoPRA11Optical, LemFri11OpticalLong},
Herrera \textit{et al.} \cite{Entag-Herrera} recently described a one-step method to 
generate entanglement between polar molecules
in the absence of dc electric fields. They 
showed that alignment-mediated entanglement in a molecular array is long-lived and
discussed applications for quantum information processing and quantum metrology.
Manipulating and controlling entanglement for molecules in external fields continue
 to be of great interest in quantum information and quantum computing. Lee \textit{et al.} experimentally explored the possibility to use rotational wavepackets created with a short laser pulse for quantum information processing~\cite{LeePRL04} (also see Ref.~\cite{ShapiroJMO05}).


Schemes for robust quantum computation with polar molecules and experimental feasibility thereof have been analyzed by Yelin, Kuznetsova, and coauthors~\cite{YelinPRA06, KuznetsovaPRA08b}. Kuznetsova~\textit{et al.} proposed a platform based on polar molecules and neutral atoms for quantum information processing~\cite{KuznetsovaPRA10, KuznetsovaQIP11}. They also proposed a scheme for cluster state generation using van der Waals and dipole-dipole interactions between atoms and molecules in an optical lattice~\cite{KuznetsovaPRA12}. The same group also proposed to use Rydberg atoms to mediate interactions between polar molecules as a tool for creating quantum gates and achieving individual addressability \cite{KuznetsovaPCCP11}. Mur-Petit~\textit{et al.} discussed the possibility to use trapped molecular ions as a qubit~\cite{MurPetit12}, and realize quantum phase gates based on polar molecules coupled to atomic ions~\cite{MurPetit13}.
De Vivie-Riedle and coworkers developed schemes for quantum computation with vibrationally excited molecules~\cite{TeschCPL01, TeschPRL02, TeschJCP04, deVivieRiedleCR07}.

\section{Stability of atomic and molecular systems in High-Frequency Super-Intense Laser Fields}
\label{Molec:stability}

Stability of atomic and molecular systems in external electric, 
magnetic and laser fields is of fundamental importance in atomic and molecular physics and has attracted considerable experimental and theoretical attention over the past decades.
It was recently shown that superintense radiation fields of sufficiently high frequency can have
large effects on the structure, stability, and ionization of atoms and molecules
\cite{kais10,kais17,kais58,kais121,kais122,kais123}. One of the most
intriguing results of Gavrila and coworkers is the possibility to stabilize multiply charged negative
ions of hydrogen by superintense laser fields \cite{kais57}. This kind of stabilization phenomena has not yet been
observed experimentally, due to the challenges of preparing and measuring such systems.
There are, however, experiments demonstrating light-induced stabilization against photoionization
when atoms are initially prepared in a Rydberg state \cite{kais56}. Recently Eichmann~\textit{et~al.} \cite{EichmannPRL13} demonstrated the high survival probability of Rydberg atoms in laser fields with intensities  above $10^{15}$ W/cm$^{2}$  experimentally, using a direct detection technique.

A classical interpretation for the stabilization, which enables an atom to bind many additional
electrons, has been given by Vorobeichik \textit{et al.} \cite{kais124}. They showed that for a sufficiently large value of the parameter
$\alpha_0=E_0/\omega^2$,  where $E_0$
and  $\omega$   are the amplitude and frequency of the laser field, the frequency associated
with the motion of the particle in the time-averaged potential,  $V_0$,  is much smaller than the laser
frequency and therefore the mean-field approach is applicable. Moiseyev and Cederbaum have shown
that the stabilization effect takes place at increasing  field strengths when,  first, the photoionization
rate decreases, and, second, the electron correlation and hence autoionization is suppressed \cite{kais126}. For
one-electron systems, Pont \textit{et al.} \cite{kais125} have shown that by increasing  $\alpha_0$, 
 the electronic eigenfunctions
of the ``dressed" potential of an atom in high-intensity laser field and the corresponding charge densities
are split into two lobes, localized around the end points of the nuclear charge, which is smeared along a
line. This phenomenon has been termed a dichotomy of the atom~\cite{kais125}.

A system of $N$ electrons and one nucleus with charge $Z$ placed in a monochromatic laser field  
$ E(t) = E_0 ({\bold e_1} \cos\omega t+{\bold e_2}  \tan\delta \sin \omega t)$
can be described within the high-frequency Floquet theory 
by the following Schr\"{o}dinger equation
\cite{kais17}:

\begin{equation}
\sum_{i=1}^N  \left( \frac{1}{2}  {\bold P_i^2 }+V_0({\bold r_i} ,\alpha_0)+\sum_{j=1}^{i-1} \frac{1}{|\bold {r_i-r_j}|} \right) \Phi = \epsilon(\alpha_0) \Phi 
\label{atom-intesen-field}
\end{equation}
where  the ``dressed'' Coulomb potential  is given by  $V_0(\bold r, \alpha_0)=-\frac{Z}{2 \pi} \int_0^{2 \pi} \frac{d\xi}{|\bold r +\alpha(\xi/\omega)|}$, with $\alpha(t) = E(t)/(m_e \omega^2)$.  This equation can be solved self-consistently in order 
to obtain the ground state energy and wave function of the system and find the critical value of  $\alpha_0$ 
for binding $N$ electrons \cite{Sergeev}. As long  as $\epsilon^{(N)}(\alpha_0)  > \epsilon^{(N-1)}(\alpha_0)  $, one of the electrons on the $N$-electron ion auto-detaches. This is always the case for multiply-charged negative ions in the absence of laser fields, therefore negative ions carrying the charge of $-2$ or more are usually unstable \cite{Scheller,Simons}. In order to determine the conditions for the stability of an multiply charged negative ions, one can  use the condition 
$D^{(N)}(\alpha_0)=\epsilon^{(N-1)}(\alpha_0)-\epsilon^{(N)}(\alpha_0)=0$.  This gives the critical value of $\alpha_0 = \tilde \alpha_0$. For values of $\alpha_0$  greater than the
$\tilde \alpha_0$, there is no auto-detachment, and the $N$-electron multiply charged negative ion supports a bound state.
Using finite size scaling method with elliptical basis functions, Wei and coworkers calculated
 all the critical parameters needed for stability of
  H$^-$ , H$^{--}$, He$^-$ , and He$^{--}$ atomic anions~\cite{Qi1,Qi2,Qi3}.


The solutions of Eq. (\ref{atom-intesen-field}) provide information on the properties of atomic systems in the presence of an intense laser field. For experimental implementations, it is necessary to consider the dynamical evolution of the system as the laser field is applied \cite{kais127}. The dynamical information can be obtained by solving the time-dependent Schr\"{o}dinger equation, as was done numerically to explore the dynamical stabilization of the ground state hydrogen atom in superintense laser pulses \cite{kais128}. In order to explore the dynamics of ion stability, one can use frequency-dependent potentials to describe the laser-driven atomic and molecular systems. Unlike the High-Frequency Floquet
theory, where the dressed potential depends on the characteristic parameter $\alpha_0=\sqrt{I}/\omega^2$, given by the ratio of the field intensity and frequency, this approach provides a time averaged potential that depends
explicitly on both $I$ and $\omega$. It was shown that this potential provides a more accurate description than the dressed potential $V_0$ \cite{kais129}, yielding a modified equation

\begin{equation}
\sum_{i=1}^N\left[ \frac{1}{2} {\bold P_i^2} +V_0(\bold r_i, \alpha_0(t)) + \frac{1}{2\omega^2} 
\sum_{n \neq 0} \frac{f_n f_{n-1}}{n^2} + \sum_{j=1}^{i-1} \frac{1}{|\bold r_i-\bold r_j|}
\right] \Phi(t)=\epsilon (\alpha_0,\omega) \Phi(t)
\end{equation}
where $V_0=-\frac{Z}{2\pi} \int_{-\pi}^{\pi}
 \frac{d\xi}{|r+\alpha(t+\xi/\omega)|}$,
 $f_n(\bold r,t)=-\frac{\partial V_n}{\partial z}$, and 
  $V_n=-\frac{Z}{2\pi} \int_{-\pi}^{\pi}
 \frac{e^{-in\xi} d\xi}{|r+\alpha(t+\xi/\omega)|}$.
The wave functions and
eigenvalues as a function of $\alpha_0(t)$ can be obtained numerically by time-dependent 
self-consistent field
methods. It was shown that for the laser frequency $\omega= 5$ eV, the laser intensity needed for
stabilization is $I=9 \cdot 10^{15}$  W/cm$^2$ and the maximal detachment energy is 1.0 eV  \cite{kais129}.

This analysis can be extended to complex atoms and molecules using
the dimensional scaling theory~\cite{Dudley123}, which provides a natural means to examine electron 
localization in super-intense laser fields. In the large-dimension limit, $D \rightarrow \infty$,
 in a suitably scaled space, electrons become
fixed along the direction of the polarized laser filed. Because of the symmetry of polarization, it is
convenient to work with cylindrical coordinates in $D$-dimensions. Herschbach and coworkers~\cite{Dudley123,Dudley124}
have shown how to generalize the Schr\"{o}dinger equation for a few elections to $D$-dimensions using cylindrical coordinates. The method is general and provides a systematic procedure to construct
large-$D$ limit effective
 Hamiltonians that are internally modified to reflect major finite-$D$ effects~\cite{Dudley125,Dudley126,tsips96}.
These functions are obtained by scaling the kinetic terms represented by generalized centrifugal potentials 
in the $D \rightarrow \infty$ limit. As applied to atoms and molecules, it was generalized to $N$-electron systems in a superintense laser field~\cite{Qi2,Qi3,Ross}. Results on the stability using the dimensional scaling are remarkably close to the three-dimensional results from both  
non-relativistic \cite{Qi2,Qi3} and relativistic~\cite{Ross}  calculations. 

The dimensional scaling theory can also be used to examine the effect of superintense laser fields on  the  binding energy of molecules, in particular, diatomic molecules~\cite{M1,M2,M3,M4}.
Results obtained from dimensional scaling with the high-frequency Floquet theory were used to
evaluate the stability of  simple diatomic molecules in the gas phase, such as H$_2^+$, H$_2$ He$_2$,
and H$_2^-$ in superintense laser fields~\cite{Qi3}. The
large-$D$ limit provides a simple model that captures the main physics of the problem, which imposes
electron localization along the polarization direction of the laser field. This localization markedly
reduces the ionization probability and can enhance chemical bonding when the laser strength
becomes sufficiently strong. 

The lack of experimental evidence for  suppression  of ionization in a high-frequency
superintense laser field looms in marked contrast with the abundance of theoretical work affirming
and elucidating stabilization. We expect that the dimensional scaling 
 method will aid the search to identify molecular systems  amenable to experimental
observation of stabilization. 

\section{Outlook}

Molecules, coming in a great variety of forms, offer a great platform for fundamental and applied research. New breakthroughs are expected if complete control is achieved over the rotational, fine-structure, hyperfine, and  translational degrees of freedom of molecules. As this article illustrates, tremendous progress has already been made towards this goal. However, there are still challenges that must be met. 

Cooling molecules to ultracold temperatures, required to harness the translational motion, remains the greatest challenge. As of now, there is still no general technique for the production of molecules at ultracold temperatures and the experiments with ultracold molecules are mostly limited to alkali metal dimers.  Achieving high-density samples of cold polyatomic molecules still represents a challenge for current experiments. It is not clear whether quantum degenerate gases of complex molecules will be stable. 

When cooled to ultracold temperatures, molecules can be loaded onto optical lattices. If done with high fidelity, this will produce ideal systems for quantum simulation and scalable quantum information processing. Trapped polar molecules are particularly interesting due to long-range interactions that couple long-lived molecular states. In order to exploit these systems for quantum simulation and quantum computing, it is necessary to develop methods for addressing molecules at individual lattice sites. In addition, it is necessary to develop robust global entanglement measures relevant to these experiments, which remains an interesting open  problem.  In order to realize a variety of many-body Hamiltonians, one needs to develop versatile techniques to manipulate the strength and symmetry of two- and many-body interactions between ultracold molecules using static and radiative fields.

 Preparing molecules in a single internal quantum state, particularly a state with high angular momentum, remains a great challenge. This limits the studies exploring the effects of internal degrees of freedom and molecular orientation or alignment on chemical dynamics. This also limits the possibility of using molecular gases for sensitive mapping of electromagnetic fields.

 The experimental work with molecules in electromagnetic fields requires the development of adequate theory. Rigorous quantum calculations of collision cross sections for molecules in external fields are highly demanding and, at present, can only be performed for collisions at low translational energies. Quantum theory of reactive scattering in external fields is a notoriously difficult problem due to the large number of states that need to be taken into account. Theoretical simulations of experiments at ultracold temperatures are impeded by the lack of numerical methods to produce intermolecular potentials with sufficient accuracy. It is necessary to develop approaches for inverting the scattering problem in order to fit intermolecular potentials, with sufficient accuracy, using the experimentally observed scattering properties of ultracold molecules.

Superintense laser fields of sufficiently high frequency might have significant effects on the structure, stability, ionization and dissociation of molecules. 
Of particular interest is the possibility of using these effects for laser-induced formation of exotic atomic and molecular systems, such as chemically bound He dimers. 
 This direction of research might open up a new field of engineering artificial atomic and molecular systems. However, there is a great need for an experimental evidence of suppression of ionization in high-frequency superintense laser fields.

In order to use molecules for practical applications, such as quantum computing, it is desirable to integrate molecular systems with solid-state devices. Such hybrid devices have been considered in a number of publications, however their implementation remains an extremely difficult task. In general, the effects of dissipation on the dynamics of molecules are not well understood. 
While it is now possible to engineer open quantum systems with controlled coupling to the environment, the possibility of using controlled dissipation as a useful resource for engineering coherent molecular states is yet to be investigated.

\clearpage
\newpage

\section{Acknowledgements}

We are grateful to (in alphabetical order) Rick Bethlem, Hans Peter B{\"u}chler, Wes Campbell, Lincoln Carr, Lorenz Cederbaum, Daniel Comparat, Bretislav Friedrich, Rosario Gonz{\'a}lez F{\'e}rez, Alexey Gorshkov, Gerrit Groenenboom, Kaden Hazzard, Dudley Herschbach, Steven Hoekstra, Emil Kirilov, Christiane Koch, Jochen K{\"u}pper, J{\"o}rn Manz, Bas van de Meerakker, Gerard Meijer, Valery Milner, Robert Moszynski, Hanns-Christoph N{\"a}gerl, Ed Narevicius, Andreas Osterwalder, Guido Pupillo, Ana Maria Rey, Peter Reynolds, Hirofumi Sakai,  Peter Schmelcher, Burkhard Schmidt, Melanie Schnell, Evgeny Shapiro, Alkwin Slenczka, Tim Softley, Henrik Stapelfeldt, Frank Stienkemeier, Steven Stolte, William Stwalley, Michael Tarbutt, Peter Toennies, Nicolas Vanhaecke, Michael Wall, Hendrik Weimer, Stefan Willitsch, Jun Ye, and Martin Zeppenfeld for comments on the manuscript.

This work was supported by NSF through a grant for the Institute for Theoretical Atomic, Molecular, and Optical Physics at Harvard University and Smithsonian Astrophysical Observatory, a  grant to the NSF CCI center ``Quantum Information for Quantum Chemistry'', and by NSERC of Canada. M.L. and R.V.K. thank KITP for hospitality.

 \bibliography{References_library}

\end{document}